\documentclass{emulateapj}

\usepackage{amssymb}
\usepackage{amsmath}
\usepackage{graphicx}
\usepackage{natbib}
\usepackage{txfonts}

\bibpunct{(}{)}{;}{a}{}{,} 

\shorttitle{On general relativistic uniformly rotating white dwarfs}
\shortauthors{Boshkayev et al.}

\def\mnras{Mon. Not. Roy. Astr. Soc.}
\def\apj{Astrophys. J.}
\def\apjl{Astrophys. J. Lett.}
\def\apjs{Astrophys. J. Suppl. Ser.}

\def\aap{Astron. Astrophys.}
\def\actaa{Acta Astronomica}

\begin{document}

\title{On general relativistic uniformly rotating white dwarfs}

\author{Kuantay Boshkayev, Jorge A. Rueda, Remo Ruffini and Ivan Siutsou}
\affil{Dipartimento di Fisica and ICRA, Sapienza Universit\`a di Roma, P.le Aldo Moro 5, I--00185 Rome, Italy}
\affil{ICRANet, P.zza della Repubblica 10, I--65122 Pescara, Italy}

\altaffiltext{}{kuantay@icra.it, jorge.rueda@icra.it, ruffini@icra.it, siutsou@icranet.org}

\begin{abstract}
The properties of uniformly rotating white dwarfs (RWDs) are analyzed within the framework of general relativity. Hartle's formalism is applied to construct the internal and external solutions to the Einstein equations. The WD matter is described by the relativistic Feynman-Metropolis-Teller equation of state which generalizes the Salpeter's one by taking into account the finite size of the nuclei, the Coulomb interactions as well as electroweak equilibrium in a self-consistent relativistic fashion. The mass $M$, radius $R$, angular momentum $J$, eccentricity $\epsilon$, and quadrupole moment $Q$ of RWDs are calculated as a function of the central density $\rho_c$ and rotation angular velocity $\Omega$. We construct the region of stability of RWDs ($J$-$M$ plane) taking into account the mass-shedding limit, inverse $\beta$-decay instability, and the boundary established by the turning-points of constant $J$ sequences which separates stable from secularly unstable configurations. We found the minimum rotation periods $\sim 0.3$, $0.5$, $0.7$ and $2.2$ seconds and maximum masses $\sim 1.500$, $1.474$, $1.467$, $1.202$ $M_\odot$ for $^{4}$He, $^{12}$C, $^{16}$O, and $^{56}$Fe WDs respectively. By using the turning-point method we found that RWDs can indeed be axisymmetrically unstable and we give the range of WD parameters where it occurs. We also construct constant rest-mass evolution tracks of RWDs at fixed chemical composition and show that, by loosing angular momentum, sub-Chandrasekhar RWDs (mass smaller than maximum static one) can experience both spin-up and spin-down epochs depending on their initial mass and rotation period while, super-Chandrasekhar RWDs (mass larger than maximum static one), only spin-up.
\end{abstract}


\maketitle

\section{Introduction}\label{sec:1}

The relevance of rotation in enhancing the maximum stable mass of a white dwarf (WD) have been discussed for many years both for uniform rotation \citep[see e.g.][]{1964ApJ...140..552J,1965PNAS...54...23A,1966ZA.....64..504R,1966MNRAS.132..305M,1989ApJS...70..661G} and differential rotation \citep[see e.g.][]{1968ApJ...151.1089O,1969ApJ...155..987O,1970A&A.....4..423T,1975ApJ...199..179D}. Newtonian gravity and post-Newtonian approximation have been mainly used to compute the structure of the star, with the exception of the work of \cite{1971Ap......7..274A}, where rotating white dwarfs (RWDs) were computed in full General Relativity (GR). From the microscopical point of view, the equation of state (EOS) of cold WD matter has been assumed to be either the one of a microscopically uniform degenerate electron fluid used by \cite{chandrasekhar31} in his classic work, or assumed to have a polytropic form.

However, as shown first by \cite{salpeter61} in the Newtonian case and then by \citep{RotC2011,RotD2011} in General Relativity (GR), a detailed description of the EOS taking into account the effects of the Coulomb interaction are essential for the determination of the maximum stable mass of non-rotating WDs. Specific microphysics of the ion-electron system forming a Coulomb lattice, together with the detail computation of the inverse $\beta$-decays and the pycnonuclear reaction rates, play a fundamental role.

A new EOS taking into account the finite size of the nucleus, the Coulomb interactions, and the electroweak equilibrium in a self-consistent relativistic fashion has been recently obtained by \cite{RotC2011}. This relativistic Feynman-Metropolis-Teller (RFMT) EOS generalizes both the \cite{chandrasekhar31} and \cite{salpeter61} works in that a full treatment of the Coulomb interaction is given through the solution of a relativistic Thomas-Fermi model. This leads to a more accurate calculation of the energy and pressure of the Wigner-Seitz cells, hence a more accurate EOS. It has been shown how the Salpeter EOS overestimates at high densities and underestimates at low densities the electron pressure. The application of this new EOS to the structure of non-rotating $^{4}$He, $^{12}$C, $^{16}$O and $^{56}$Fe was recently done in \citep{RotD2011}. The new mass-radius relations generalize the works of \cite{chandrasekhar31} and \cite{hamada61}; smaller maximum masses and a larger minimum radii are obtained. Both GR and inverse $\beta$-decay can be relevant for the instability of non-rotating WDs depending on the nuclear composition, as we can see from Table \ref{tab:mcrit}, which summarizes some results of \cite{RotD2011}.

\begin{table}[h]
\centering
\begin{tabular}{c c c c}
Composition & $\rho^{J=0}_{\rm crit}$ (g/cm$^3$) & Instability & $M^{J=0}_{max}/M_\odot$\\
\colrule $^{4}$He &  $1.56\times 10^{10}$&GR & 1.40906 \\
$^{12}$C & $2.12\times 10^{10}$ & GR& 1.38603 \\
$^{16}$O & $1.94\times 10^{10}$& inverse $\beta$-decay & 1.38024 \\
$^{56}$Fe & $1.18\times 10^9$ & inverse $\beta$-decay& 1.10618\\
\botrule
\end{tabular}
\caption{Critical density and mass for the gravitational collapse of non-rotating $^4$He, $^{12}$C, $^{16}$O and $^{56}$Fe WDs in GR obtained by \cite{RotD2011}, based on the RFMT EOS \cite{RotC2011}. We indicate in the third column if the critical density is due either to inverse $\beta$-decay or to general relativistic effects.}
\label{tab:mcrit}
\end{table}

We here extend the previous results of \cite{RotD2011} for uniformly RWDs at zero temperatures obeying the RFMT EOS. We use the Hartle's approach \citep{H1967} to solve the Einstein equations accurately up to second order approximation in the angular velocity of the star. We calculate the mass $M$, equatorial $R_{eq}$ and polar $R_p$ radii, angular momentum $J$, eccentricity $\epsilon$, and quadrupole moment $Q$, as a function of the central density $\rho_c$ and rotation angular velocity $\Omega$ of the WD. We construct also RWD models for the Chandrasekhar and Salpeter EOS and compare and contrast the differences with the RFMT ones.

We analyze in detail the stability of RWDs both from the microscopic and macroscopic point of view in Sec.~\ref{sec:3}. Besides the inverse $\beta$-decay instability, we also study the limits to the matter density imposed by zero-temperature pycnonuclear fusion reactions using up-to-date theoretical models \citep{pycyak2005,pycyak2006}. The mass-shedding limit as well as the secular axisymmetric instability boundary are calculated.

The general structure and stability boundaries of $^4$He, $^{12}$C, $^{16}$O and $^{56}$Fe WDs are discussed in in Sec.~\ref{sec:4}. From the maximally rotating models (mass-shedding sequence), we calculate in Sec.~\ref{sec:5} the maximum mass of uniformly rotating $^4$He, $^{12}$C, $^{16}$O and $^{56}$Fe WDs for the Chandrasekhar, Salpeter, and RFMT EOS, and compare the results with the existing values in the literature. We calculate the minimum(maximum) rotation period(frequency) of a RWD for the above nuclear compositions, taking into account both inverse $\beta$-decay and pycnonuclear restrictions to the density; see Sec.~\ref{sec:6}.

We discuss in Sec.~\ref{sec:7} the axisymmetric instabilities found in this work. A comparison of Newtonian and general relativistic WDs presented in App.~\ref{app:3} show that this is indeed a general relativistic effect. Furthermore, we estimate in App.~\ref{app:4} the accuracy of the ``slow'' rotation approximation (power-series solutions up to order $\Omega^2$) for the determination of the maximally rotating sequence of WDs. In this line, we calculate the rotation to gravitational energy ratio and the deviations from spherical symmetry.

In addition, we construct in Sec.~\ref{sec:8} constant rest-mass evolution tracks of RWDs at fixed chemical composition and show that RWDs may experience both spin-up and spin-down epochs while loosing angular momentum, depending on their initial mass and rotation period.

Finally, in Sec.~\ref{sec:9} we outline some astrophysical implications of the results presented in this work, which we summarized in Sec.~\ref{sec:10}.

\section{Spacetime geometry and Hartle's formalism}\label{sec:2}

\cite{H1967} described for the first time the structure of rotating objects approximately up to second order terms in the angular velocity of the star $\Omega$, within GR. In this ``slow'' rotation approximation, the solution of the Einstein equations in the exterior vacuum can be written in analytic closed form in terms of the total mass $M$, angular momentum $J$ and quadrupole moment $Q$ of the star (see App.~\ref{app:1}). The interior metric is constructed by solving numerically a system of ordinary differential equations for the perturbation functions \citep[see][for details]{H1967,HT1968}. 

The spacetime geometry up to order $\Omega^2$, with an appropriate choice of coordinates is, in geometrical units $c=G=1$, described by \citep{H1967}
\begin{eqnarray}\label{eq:h1}
&ds^{2}& = \left\{ e^{\nu(r)}[1+2 h_0(r)+2 h_2(r)P_2(\cos\theta)] -\omega^2 r^2 \sin^2\theta\right\}dt^{2} \nonumber \\ 
&+& 2 \omega r^2 \sin^2\theta dt d\phi -e^{\lambda(r)}\left[1+2 \frac{m_0(r)+m_2(r)P_2(\cos\theta)}{r-M^{J=0}(r)}\right]dr^2 \nonumber \\ &-&r^2\left[1+2k_2(r)P_2(\cos\theta)\right](d\theta^2+\sin^2\theta d\phi^2)\, ,
\end{eqnarray}
where $P_2(\cos\theta)$ is the Legendre polynomial of second order, $e^{\nu(r)}$ and $e^{\lambda(r)}=[1-2 M^{J=0}(r)/r]^{-1}$, and $M^{J=0}(r)$ are the metric functions and mass of the corresponding static (non-rotating) solution with the same central density as the rotating one. The angular velocity of local inertial frames $\omega(r)$, proportional to $\Omega$, as well as the functions $h_0$, $h_2$, $m_0$, $m_2$, $k_2$, proportional to $\Omega^2$, must be calculated from the Einstein equations \citep[see][for details]{H1967,HT1968}; their analytic expressions in the vacuum case can be found in App.~\ref{app:1}.

The parameters $M$, $J$ and $Q$, are then obtained for a given EOS from the matching procedure between the internal and external solutions at the surface of the rotating star. The total mass is defined by $M=M^{J\neq0}=M^{J=0}+\delta M$, where $M^{J=0}$ is the mass of a static (non-rotating) WD with the same central density as $M^{J\neq0}$, and $\delta M$ is the contribution to the mass due to rotation.
%

\section{Limits on the stability of rotating white dwarfs}\label{sec:3}

\subsection{The mass-shedding limit}\label{sec:3a}

The velocity of particles on the equator of the star cannot exceed the Keplerian velocity of ``free'' particles, computed at the same location. In this limit, particles on the star's surface keep bound to the star only due to a balance between gravity and centrifugal forces. The evolution of a star rotating at this Keplerian rate is accompanied by loss of mass, becoming thus unstable \citep[see e.g.][for details]{stergioulas}. A procedure to obtain the maximum possible angular velocity of the star before reaching this limit was developed e.g. by \cite{Friedman1986}. However, in practice, it is less complicated to compute the mass-shedding (or Keplerian) angular velocity of a rotating star, $\Omega^{J \neq 0}_{K}$, by calculating the orbital angular velocity of a test particle in the external field of the star and corotating with it at its equatorial radius, $r=R_{eq}$.

For the Hartle-Thorne external solution, the Keplerian angular velocity can be written as (see e.g.~\cite{2008AcA....58....1T,Bini2011} and App.~\ref{app:1b}, for details)
\begin{equation}\label{eq:omegaK}
\Omega^{J \neq 0}_{K}=\sqrt{G \frac{M}{R_{eq}^3}}\left[1- j F_{1}(R_{eq})+j^2F_{2}(R_{eq})+q F_{3}(R_{eq})\right],
\end{equation}
where $j=c J/(G M^2)$ and $q=c^4 Q/(G^2 M^3)$ are the dimensionless angular momentum and quadrupole moment, and the functions $F_i(r)$ are defined in App.~\ref{app:1b}. Thus, the numerical value of $\Omega^{J \neq 0}_{K}$ can be computed by gradually increasing the value of the angular velocity of the star, $\Omega$, until it reaches the value $\Omega^{J \neq 0}_{K}$ expressed by Eq.~(\ref{eq:omegaK}).

It is important to analyze the issue of the accuracy of the slow rotation approximations, e.g.~accurate up to second order in the rotation expansion parameter, for the description of maximally rotating stars as WDs and neutron stars (NSs). We have performed in App.~\ref{app:4} a scrutiny of the actual physical request made by the slow rotation regime. Based on this analysis, we have checked that the accuracy of the slow rotation approximation increases with the density of the WD, and that the mass-shedding (Keplerian) sequence of RWDs can be accurately described by the $\Omega^2$ approximation within an error smaller than the one found for rapidly rotating NSs, $\lesssim 6$\%.

\subsection{The turning-point criterion and secular axisymmetric instability}\label{sec:3b}

In a sequence of increasing central density the mass of non-rotating star is limited by the first maximum of the $M$-$\rho_c$ curve, i.e. the turning-point given by the maximum mass, $\partial M/\partial \rho_c = 0$, marks the secular instability point and it coincides also with the dynamical instability point if the perturbation obeys the same EOS as of the equilibrium configuration \citep[see e.g.][for details]{shapirobook}. The situations is, however, much more complicated in the case of rotating stars; the determination of axisymmetric dynamical instability points implies to find the perturbed solutions with zero frequency modes, that is, perturbed configurations whose energy (mass) is the same as the unperturbed (equilibrium) one, at second order. However, \cite{1988ApJ...325..722F} formulated, based on the works of \cite{1981ApJ...249..254S,1982ApJ...257..847S}, a turning-point method to locate the points where secular instability sets in for uniformly rotating relativistic stars: along a sequence of rotating stars with fixed angular momentum and increasing central density, the onset of secular axisymmetric instability is given by
\begin{equation}\label{eq:Jcons}
\left(\frac{\partial M(\rho_c, J)}{\partial \rho_c}\right)_{J}=0\, .
\end{equation}

Thus, configurations on the right-side of the maximum mass of a $J$-constant sequence are secularly unstable. After the secular instability sets in, the configuration evolves quasi-stationarily until it reaches a point of dynamical instability where gravitational collapse should take place \citep[see][and references therein]{stergioulas}. The secular instability boundary thus separates stable from unstable stars. It is worth stressing here that the turning-point of a constant $J$ sequence is a sufficient but not a necessary condition for secular instability and therefore it establishes an absolute upper bound for the mass (at constant $J$). We construct the boundary given by the turning-points of constant angular momentum sequences as given by Eq.~(\ref{eq:Jcons}). The question whether dynamically unstable RWDs can exist or not on the left-side of the turning-point boundary remains an interesting problem and deserves further attention in view of the very recent results obtained by \cite{2011MNRAS.416L...1T} for some models of rapidly rotating NSs.

\subsection{Inverse $\beta$-decay instability}\label{sec:3c}

It is known that a WD might become unstable against the inverse $\beta$-decay process $(Z,A)\to (Z-1,A)$ through the capture of energetic electrons. In order to trigger such a process, the electron Fermi energy (with the rest-mass subtracted off) must be larger than the mass difference between the initial $(Z,A)$ and final $(Z-1,A)$ nucleus. We denote this threshold energy as $\epsilon^\beta_Z$. Usually it is satisfied $\epsilon^\beta_{Z-1} < \epsilon^\beta_Z$ and therefore the initial nucleus undergoes two successive decays, i.e. $(Z,A)\to (Z-1,A)\to (Z-2,A)$ (see e.g.~\cite{salpeter61,shapirobook}). Some of the possible decay channels in WDs with the corresponding known experimental threshold energies $\epsilon^\beta_Z$ are listed in Table \ref{tab:betadecay}. The electrons in the WD may eventually reach the threshold energy to trigger a given decay at some critical density $\rho^{\beta}_{\rm crit}$. Since the electrons are responsible for the internal pressure of the WD, configurations with $\rho > \rho^{\beta}_{\rm crit}$ become unstable due to the softening of the EOS as a result of the electron capture process (see \cite{harrison58,salpeter61} for details). In Table \ref{tab:betadecay}, correspondingly to each threshold energy $\epsilon^\beta_Z$, the critical density $\rho^{\beta}_{\rm crit}$ given by the RFMT EOS is shown; see \cite{RotD2011} for details.
\begin{table}
\centering
\begin{tabular}{c c c c}
Decay & $\epsilon^\beta_Z$ (MeV) & $\rho^{\beta}_{\rm crit}$ (g/cm$^3$)\\
\colrule
$^{4}$He $\to ^3$ H + $n \to 4 n$ & 20.596 & $1.39\times 10^{11}$ \\
$^{12}$C $\to ^{12}$B $\to ^{12}$Be & 13.370 & $3.97\times 10^{10}$ \\
$^{16}$O $\to ^{16}$N $\to ^{16}$C& 10.419 & $1.94\times 10^{10}$ \\
$^{56}$Fe $\to ^{56}$Mn $\to ^{56}$Cr& 3.695 & $1.18\times 10^{9}$ \\
\botrule
\end{tabular}
\caption{Onset for the inverse $\beta$-decay of $^{4}$He, $^{12}$C, $^{16}$O and $^{56}$Fe. The experimental values of the threshold energies $\epsilon^\beta_Z$ have been taken from Table 1 of
\cite{audi03}; see also \citep{1977ADNDT..19..175W,shapirobook}. The corresponding critical density $\rho^{\beta}_{\rm crit}$ are for the RFMT EOS \citep[see][]{RotD2011}}.\label{tab:betadecay}
\end{table}

\subsection{Pycnonuclear fusion reactions}\label{sec:3d}

In our WD model, we assume a unique nuclear composition $(Z,A)$ throughout the star. We have just seen that inverse $\beta$-decay imposes a limit to the density of the WD over which the current nuclear composition changes from $(Z,A)$ to $(Z-1,A)$. There is an additional limit to the nuclear composition of a WD. Nuclear reactions proceed with the overcoming of the Coulomb barrier by the nuclei in the lattice. In the present case of zero temperatures $T=0$, the Coulomb barrier can be overcome because the zero-point energy of the nuclei \citep[see e.g.][]{shapirobook}
\begin{equation}\label{eq:Ep}
E_p=\hbar \omega_p\, ,\qquad \omega_p=\sqrt{\frac{4 \pi e^2 Z^2 \rho}{A^2 M_u^2}}\, ,
\end{equation}
where $e$ is the fundamental charge and $M_u=1.6605\times 10^{-24}$ g is the atomic mass unit.

Based on the pycnonuclear rates computed by \cite{zeldovich1958} and \cite{cameron1959}, \cite{salpeter61} estimated that in a time of $0.1$ Myr, $^1$H is converted into $^4$He at $\rho\sim 5\times 10^4$ g cm$^{-3}$, $^4$He into $^{12}$C at $\rho\sim 8\times 10^8$ g cm$^{-3}$, and $^{12}$C into $^{24}$Mg at $\rho\sim 6\times 10^9$ g cm$^{-3}$. The threshold density for the pycnonuclear fusion of $^{16}$O occurs, for the same reaction time $0.1$ Myr, at $\rho\sim 3\times 10^{11}$ g cm$^{-3}$, and for $10$ Gyr at $\sim 10^{11}$ g cm$^{-3}$. These densities are much higher that the corresponding density for inverse $\beta$-decay of $^{16}$O, $\rho\sim 1.9\times 10^{10}$ g cm$^{-3}$ (see Table \ref{tab:betadecay}). The same argument applies to heavier compositions e.g. $^{56}$Fe; so pycnonuclear reactions are not important for heavier than $^{12}$C in WDs. 

It is important to analyze the case of $^4$He WDs in detail. At densities $\rho_{pyc}\sim 8\times 10^8$ g cm$^{-3}$ a $^4$He WD should have a mass $M \sim 1.35 M_\odot$ \citep[see e.g.~Fig.~3 in][]{RotD2011}. However, the mass of $^4$He WDs is constrained to lower values from their previous thermonuclear evolution: a cold star with mass $>0.5 M_\odot$ have already burned an appreciable part of its Helium content at earlier stages. Thus, WDs of $M>0.5 M_\odot$ with $^4$He cores are very unlikely \citep[see][for details]{hamada61}. It should be stressed that $^4$He WDs with $M\lesssim 0.5 M_\odot$ have central densities $\rho\sim 10^6$ g cm$^{-3}$ \citep{RotD2011} and at such densities pycnonuclear reaction times are longer than 10 Gyr, hence unimportant. However, we construct in this work $^4$He RWDs configurations all the way up to their inverse $\beta$-decay limiting density for the sake of completeness, keeping in mind that the theoretical $^4$He WDs configurations with $M\gtrsim 0.5 M_\odot$ could actually not be present in any astrophysical system.


From the above discussion we conclude that pycnonuclear reactions can be relevant only for $^{12}$C WDs. It is important to stress here that the reason for which the pycnonuclear reaction time, $\tau^{\rm C+C}_{pyc}$, determines the lifetime of a $^{12}$C WD is that reaction times $\tau^{\rm C+C}_{pyc}<10$ Gyr are achieved at densities $\sim 10^{10}$ g cm$^{-3}$, lower than the inverse $\beta$ decay threshold density of $^{24}$Mg, $^{24}$Mg$\to ^{24}$Na$\to ^{24}$Ne, $\rho\sim 3.2\times 10^9$ g cm$^{-3}$ \citep[see e.g.][]{salpeter61,shapirobook}. Thus, the pycnonuclear $^{12}$C+$^{12}$C fusion produces unstable $^{24}$Mg that almost instantaneously decay owing to electron captures, and so the WD becomes unstable as we discussed in Subsec.~\ref{sec:3a}.

However, the pycnonuclear reaction rates are not known with precision due to theoretical and experimental uncertainties. \cite{hamada61} had already pointed out in their work that the above pycnonuclear density thresholds are reliable only within a factor 3 or 4. The uncertainties are related to the precise knowledge of the Coulomb tunneling in the high density low temperature regime relevant to astrophysical systems, e.g. WDs and NSs, as well as with the precise structure of the lattice; impurities, crystal imperfections, as well as the inhomogeneities of the local electron distribution and finite temperature effects, also affect the reaction rates. The energies for which the so-called astrophysical $S$-factors are known from experiments are larger with respect to the energies found in WD and NS crusts, and therefore the value of the $S$-factors have to be obtained theoretically from the extrapolation of experimental values using appropriate nuclear models, which at the same time are poorly constrained. A detailed comparison between the different theoretical methods and approximations used for the computation of the pycnonuclear reaction rates can be found in \citep{pycyak2005,pycyak2006}. 

\begin{figure}
\centering\includegraphics[width=\columnwidth,clip]{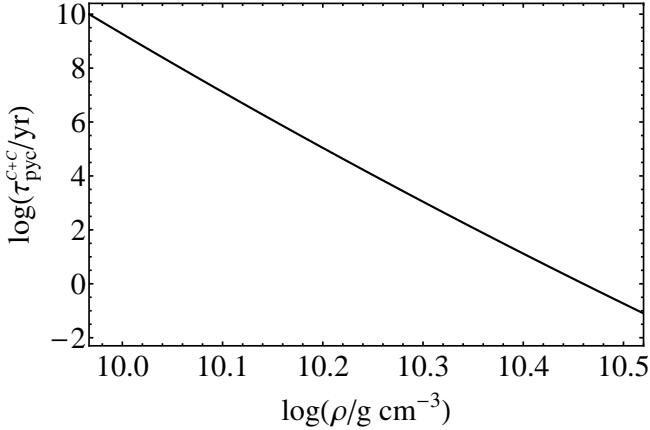}
\caption{Pycnonuclear reaction times at zero temperature for C+C fusion as a function of the density.}\label{fig:taupyc}
\end{figure}

The $S$-factors have been computed in \citep{pycyak2005,pycyak2006} using up-to-date nuclear models. Following these works, we have computed the pycnonuclear reaction times for C+C fusion as a function of the density as given by Eq.~(\ref{eq:taupyc}), $\tau^{\rm C+C}_{pyc}$, which we show in Fig.~\ref{fig:taupyc}; we refer to App.~\ref{app:2} for details. 

We obtain that for $\tau^{\rm C+C}_{pyc}=10$ Gyr, $\rho_{pyc}\sim 9.26\times 10^9$ g cm$^{-3}$ while, for $\tau^{\rm C+C}_{pyc}=0.1$ Myr, $\rho_{pyc}\sim 1.59\times 10^{10}$ g cm$^{-3}$, to be compared with the value $\rho\sim 6\times 10^{9}$ g cm$^{-3}$ estimated by \cite{salpeter61}. In order to compare the threshold densities for inverse $\beta$-decay and pycnonuclear fusion rates, we shall indicate in our mass-density and mass-radius relations the above two density values corresponding to these two lifetimes. It is important to stress that the computation of the pycnonuclear reactions rates is subjected to theoretical and experimental uncertainties \citep[see][for details]{pycyak2005}. For instance, \cite{hamada61} stated that these pycnonuclear critical densities are reliable within a factor 3 or 4. If three times larger, the above value of $\rho_{pyc}$ for $\tau^{\rm C+C}_{pyc}=0.1$ Myr becomes $\rho_{pyc}\sim 4.8\times 10^{10}$ g cm$^{-3}$, larger than the inverse $\beta$-decay threshold density $\rho^{\rm C}_{\beta}\sim 3.97\times 10^{10}$ g cm$^{-3}$ (see Table \ref{tab:betadecay}). As we will see in Sec.~\ref{sec:7}, the turning-point construction leads to an axisymmetric instability boundary in the density range $\rho^{{\rm C}, J=0}_{\rm crit}=2.12\times 10^{10}<\rho<\rho^{\rm C}_{\beta}$ g cm$^{-3}$ in a specific range of angular velocities. This range of densities is particularly close to the above values of $\rho_{pyc}$ which suggests a possible competition between different instabilities at high densities.

\section{WD structure and stability boundaries}\label{sec:4}

The structure of uniformly RWDs have been studied by several authors \citep[see e.g.][]{1964ApJ...140..552J,1965PNAS...54...23A,1966ZA.....64..504R,1966MNRAS.132..305M,1989ApJS...70..661G}. The issue of the stability of both uniformly and differentially rotating WDs has been studied as well \citep[see e.g.][]{1968ApJ...151.1089O,1969ApJ...155..987O,1970A&A.....4..423T,1975ApJ...199..179D}. All the above computations were carried out within Newtonian gravity or at the post-Newtonian approximation. The EOS of cold WD matter has been assumed to be either the one of a microscopically uniform degenerate electron fluid, which we refer hereafter as Chandrasekhar EOS \citep{chandrasekhar31}, or assuming a polytropic EOS. However, microscopic screening caused by Coulomb interactions as well as the process of inverse $\beta$-decay of the composing nuclei cannot be properly studied within such EOS \citep[see][for details]{RotC2011,RotD2011}. 

The role of general relativistic effects, shown in \cite{RotD2011}, has been neglected in all the above precedent literature. The only exception to this rule is, up to our knowledge, the work of \cite{1971Ap......7..274A}, who investigated uniformly RWDs for the Chandrasekhar EOS within GR. They use an $\Omega^2$ approximation following a method developed by \cite{sedrakyan1968}, independently of the work of \cite{H1967}. A detailed comparison of our results with the ones of \cite{1971Ap......7..274A} can be found in App.~\ref{app:3}.

\begin{figure*}
\includegraphics[width=0.48\hsize,clip]{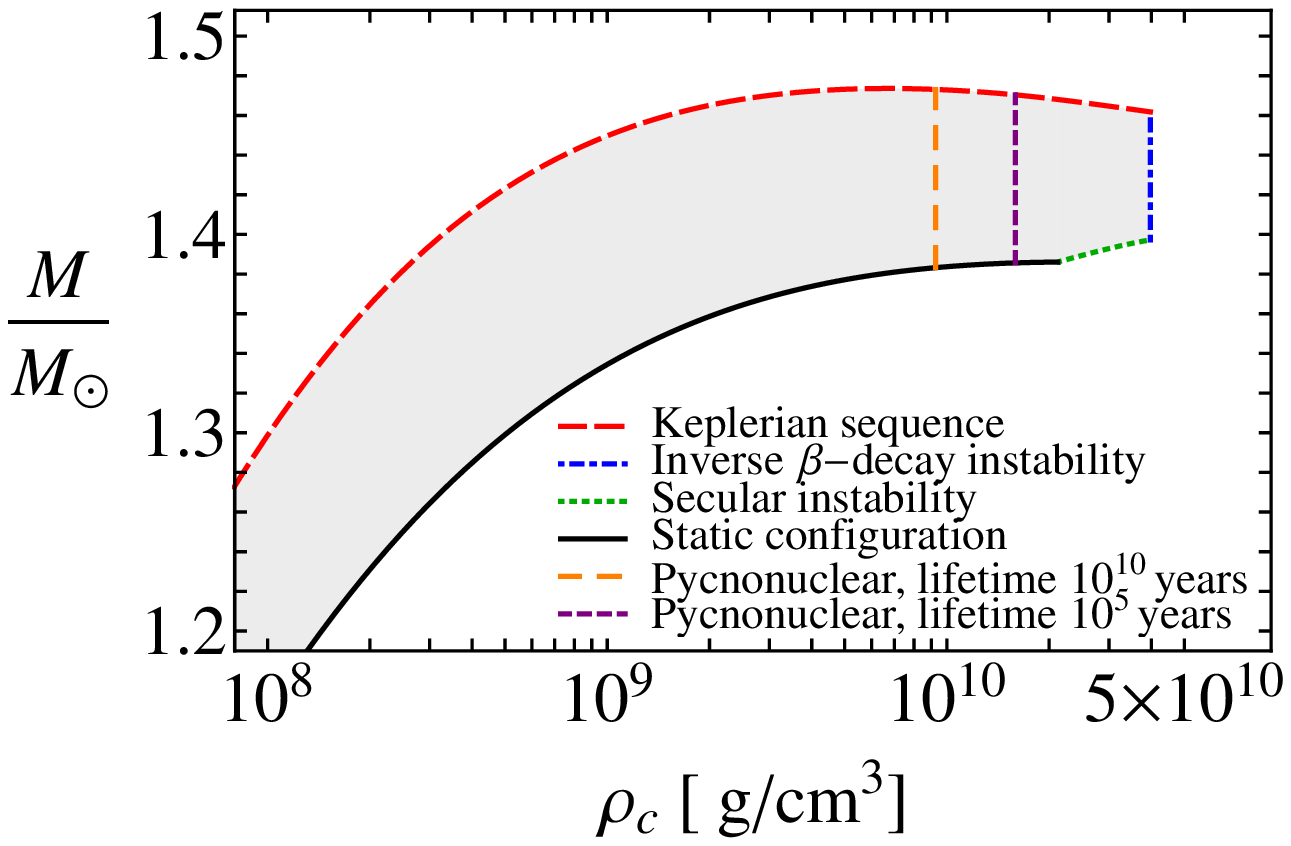} \includegraphics[width=0.48\hsize,clip]{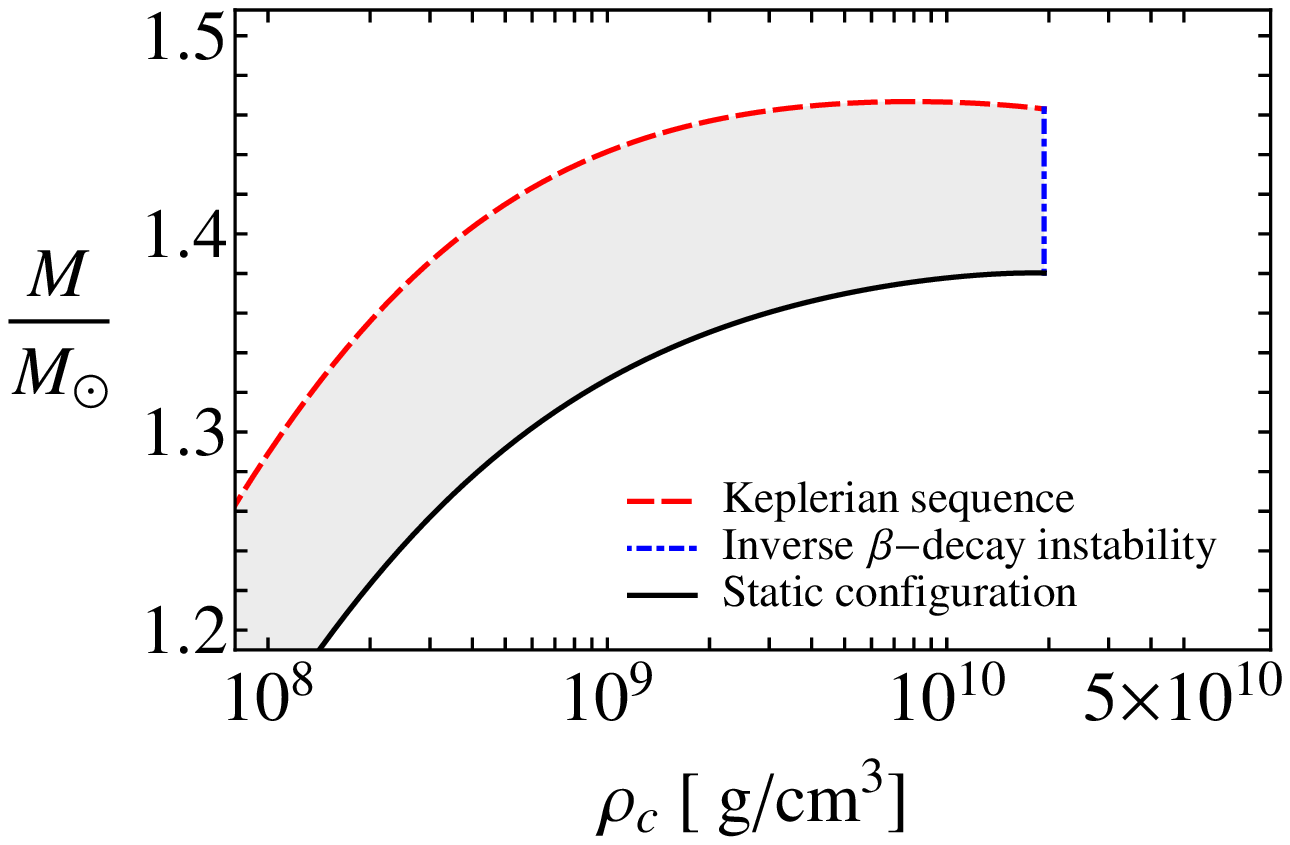}
\caption{Mass in solar masses versus the central density for $^{12}$C (left panel) and for $^{16}$O (right panel) WDs. The solid curve corresponds to the mass of non-rotating WDs, the Keplerian sequence is the red thick dashed curve, the blue thick dotted-dashed curve is the inverse $\beta$ instability boundary, and the green thick solid curve is the axisymmetric instability boundary. The orange and purple dashed boundaries correspond to the pycnonuclear densities for reaction times $\tau_{pyc}=10$ Gyr and 0.1 Myr, respectively. All rotating stable WDs are in the shaded region.}\label{fig:MrhoCO}
\end{figure*}

\begin{figure*}
\includegraphics[width=0.48\hsize,clip]{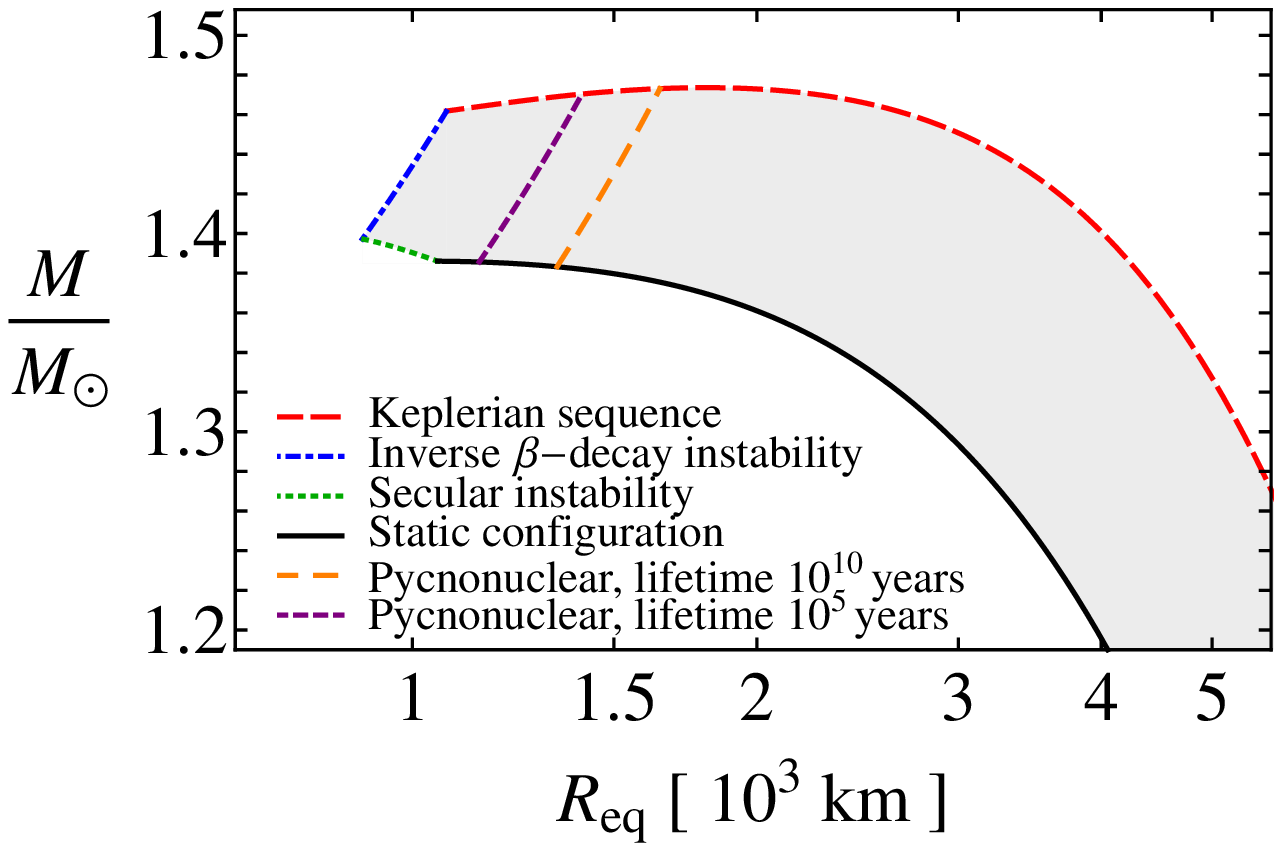} \includegraphics[width=0.48\hsize,clip]{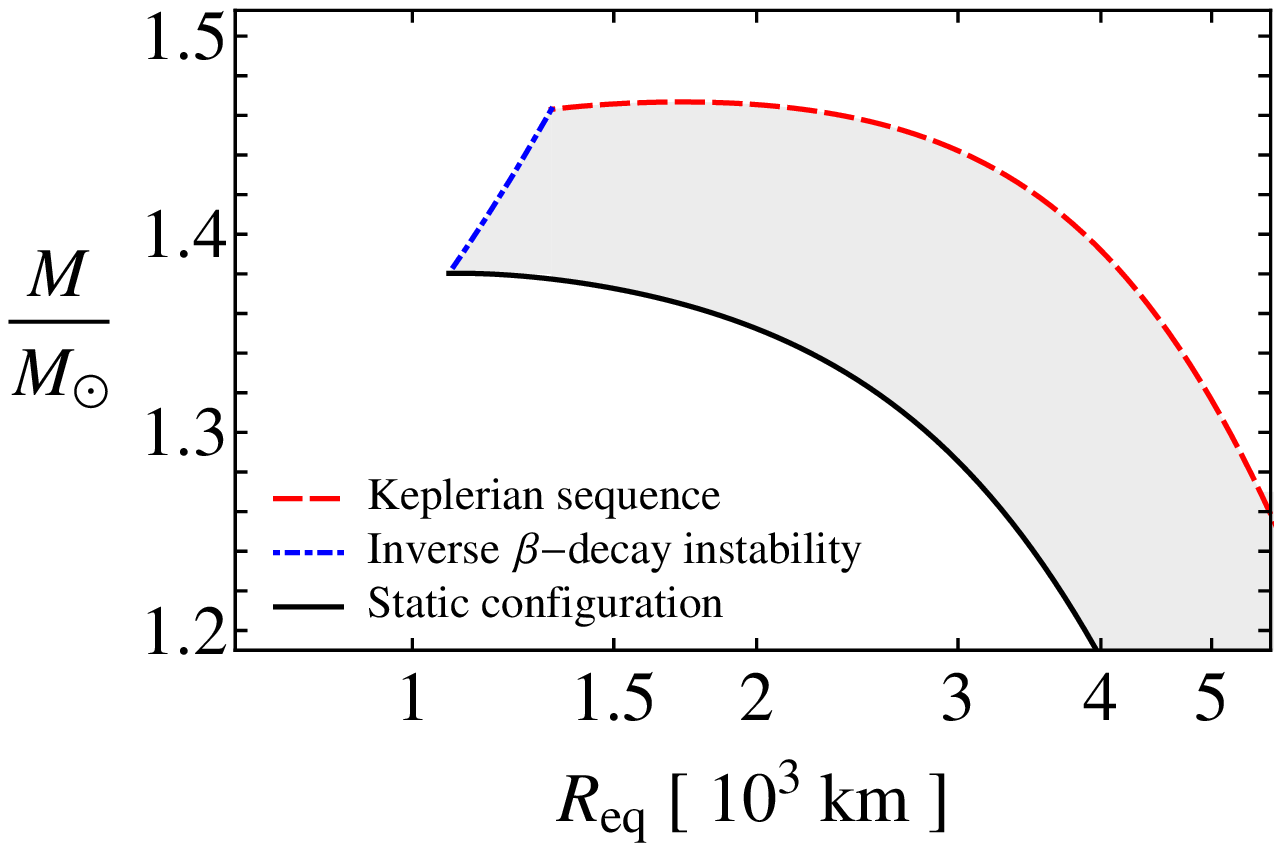}
\caption{Mass in solar masses versus the equatorial radius in units of $10^3$ km for $^{12}$C (left panel) and for $^{16}$O (right panel) WDs. The left and right panels show the configurations for the same range of central densities of the corresponding panels of Fig.~\ref{fig:MrhoCO}.}\label{fig:MRCO}
\end{figure*}

In Figs.~\ref{fig:MrhoCO}--\ref{fig:MRCO} we show the mass-central density relation and the mass-radius relation of general relativistic rotating $^{12}$C and $^{16}$O WDs. We explicitly show the boundaries of mass-shedding, secular axisymmetric instability, inverse $\beta$-decay, and pycnonculear reactions.

\begin{table*}
\centering
\begin{tabular}{c c c c c c c c c c c c}
Composition & $\rho_{M^{J\neq0}_{max}}$  & $k$ & $M^{J=0}_{max}/M_{\odot}$ & $R_{M^{J=0}_{max}}$ & $P_{min}$ &  $R_{p}^{P_{min}}$ & $R_{eq}^{P_{min}}$ & $(T/|W|)^{P_{min}}$ & $\epsilon^{P_{min}}$ & $j^{P_{min}}$ & $q^{P_{min}}$\\
\colrule
$^4$He    & 5.46$\times$10$^9$ & 1.0646 & 1.40906 & 1163 & 0.284 & 564  & 736  & 0.0163 & 0.642 & 1.004 & 526\\
$^{12}$C  & 6.95$\times$10$^9$ & 1.0632 & 1.38603 & 1051 & 0.501 & 817  & 1071 & 0.0181 & 0.647 & 1.287 & 1330\\
$^{16}$O  & 7.68$\times$10$^9$ & 1.0626 & 1.38024 & 1076 & 0.687 & 1005 & 1323 & 0.0194 & 0.651 & 1.489 & 2263\\
$^{56}$Fe & 1.18$\times$10$^9$ & 1.0864 & 1.10618 & 2181 & 2.195 & 2000 & 2686 & 0.0278 & 0.667 & 2.879 & 23702\\
\botrule
\end{tabular}
\caption{Properties of uniformly rotating general relativistic $^4$He, $^{12}$C, $^{16}$O and $^{56}$Fe WDs: 
$\rho_{M_{max}^{J\neq 0}}$ is the central density in g cm$^{-3}$ corresponding to the rotating maximum mass $M_{max}^{J\neq 0}$; $k$ is the dimensionless factor used to express the rotating maximum mass $M_{max}^{J\neq 0}$ as a function of the non-rotating maximum mass $M^{J=0}_{max}$ of WDs, in solar masses, obtained in \cite{RotD2011}, as defined in Eq.~(\ref{eq:rotMmax}); the corresponding minimum radius is $R_{M^{J=0}_{max}}$, in km; $P_{min}$ is the minimum rotation period in seconds. We recall that the configuration with $P_{min}$ is obtained for a WD rotating at the mass-shedding limit and with central density equal to the critical density for inverse $\beta$-decay (see Table \ref{tab:betadecay} and the right panel of Fig.~\ref{fig:consOmJ}). The polar $R_{p}^{P_{min}}$ and equatorial $R_{eq}^{P_{min}}$ radii of the configuration with $P_{min}$ are also given in km. The quantity $(T/|W|)^{P_{min}}$ is the ratio between the kinetic and binding energies, the parameter $\epsilon^{P_{min}}$ is the eccentricity of the star, rotating at $P_{min}$. Finally, $j^{P_{min}}$ and $q^{P_{min}}$ are the dimensionless angular momentum and quadrupole moment of WDs, respectively.}\label{tab:concl}
\end{table*}

\begin{figure*}
\includegraphics[width=0.48\hsize,clip]{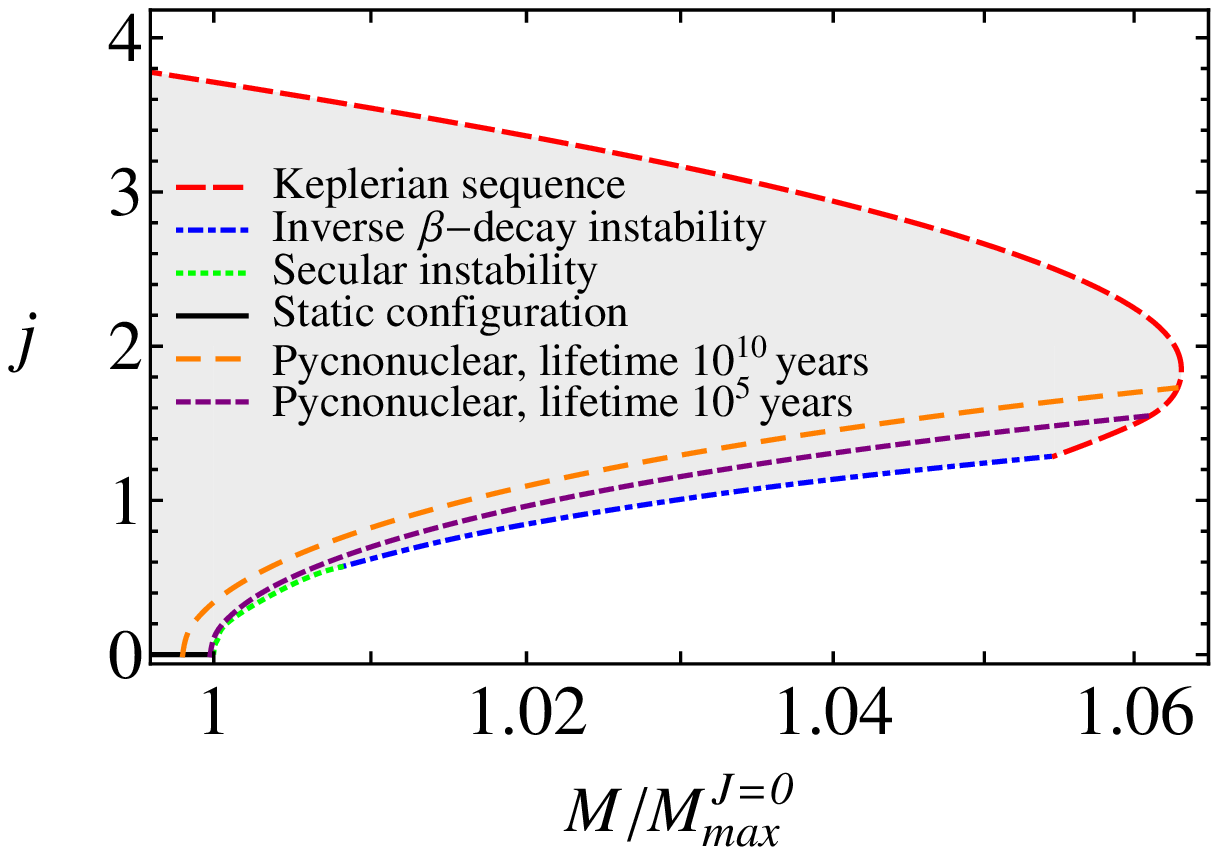} \includegraphics[width=0.48\hsize,clip]{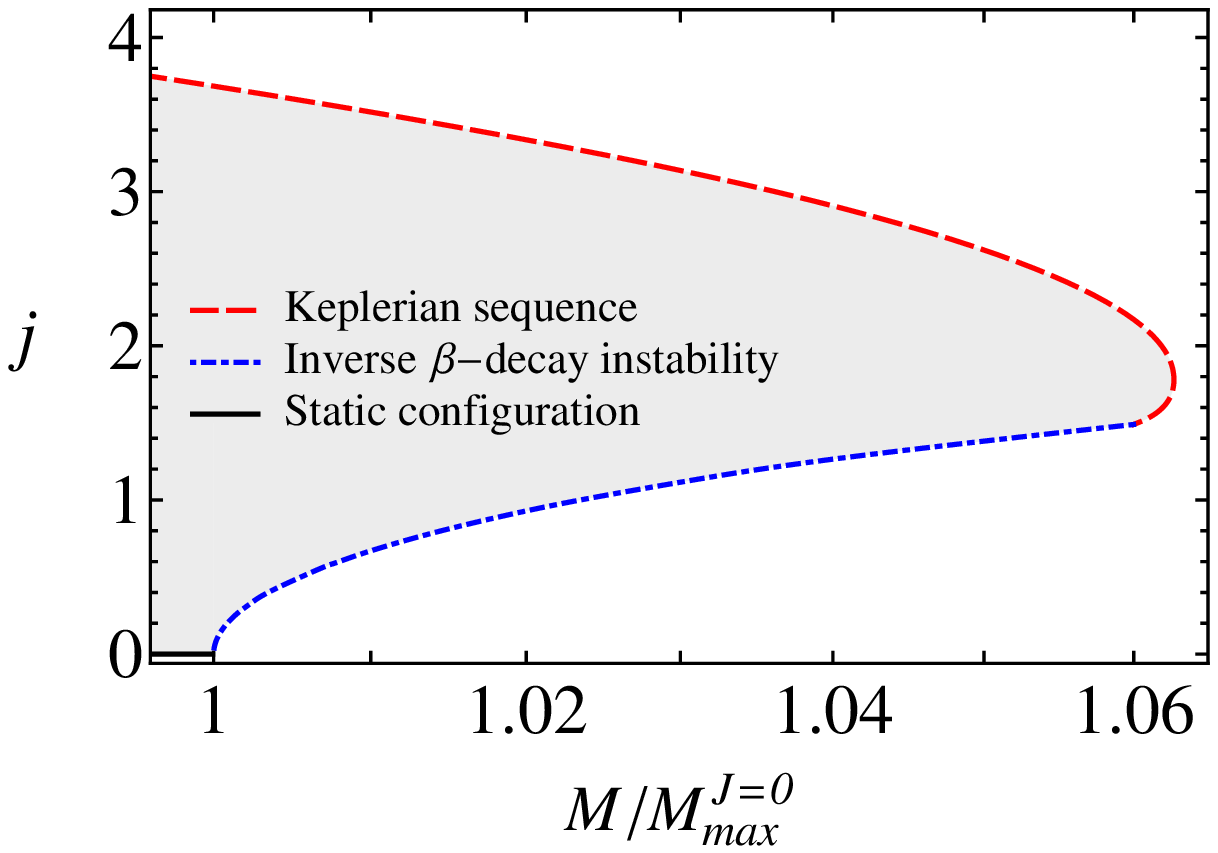}
\caption{Dimensionless angular momentum $j\equiv c J/(G M^2)$ versus the mass of rotating $^{12}$C (left panel) and $^{16}$O (right panel) WDs, normalized to the maximum non-rotating mass. All rotating stable WDs are in the shaded region.}\label{fig:JMCO}
\end{figure*}

Turning now to the rotation properties, in Fig.~\ref{fig:JMCO} we show the $J$-$M$ plane especially focusing on RWDs with masses larger than the maximum non-rotating mass, hereafter Super-Chandrasekhar WDs (SCWDs). It becomes clear from this diagram that SCWDs can be stable only by virtue of their non-zero angular momentum: the lower-half of the stability line of Fig.~\ref{fig:JMCO}, from $J=0$ at $M/M^{J=0}_{max}$ all the way up to the value of $J$ at $M^{J\neq0}_{max} \sim 1.06 M^{J=0}_{max}$, determines the critical(minimum) angular momentum under which a SCWDs becomes unstable. The upper half of the stability line determines, instead, the maximum angular momentum that SCWDs can have.

\section{The maximum mass}\label{sec:5}

The maximum masses of rotating WDs belongs to the Keplerian sequence (see Figs.~\ref{fig:MrhoCO}--\ref{fig:JMCO}) and it can be expressed as
\begin{equation}\label{eq:rotMmax}
M_{max}^{J\neq 0}=k\,M_{max}^{J=0}\, ,
\end{equation}
where $M_{max}^{J=0}$ is the maximum stable mass of non-rotating WDs and $k$ is a numerical factor that depends on the chemical composition, see Table \ref{tab:concl} for details. For $^{4}$He, $^{12}$C, $^{16}$O, and $^{56}$Fe RWDs, we found $M_{max}^{J\neq 0} \sim 1.500$, $1.474$, $1.467$, $1.202$ $M_\odot$, respectively. 

In Table \ref{tab:mrotmax} we compare the properties of the configuration with maximum mass using different EOS, namely Chandrasekhar $\mu=2$ \citep[see e.g.]{2011IJMPE..20..136B}, Salpeter, and RFMT EOS. A comparison with classical results obtained with different treatments and EOS can be found in App.~\ref{app:3}.

\begin{table*}
\centering
{\begin{tabular}{@{}ccccccc@{}}
Nuclear Composition    & EOS        & $\rho_{M^{J\neq0}_{max}}$ (g/cm$^3$) & $R^{M^{J\neq0}_{max}}_{p}$ (km) & $R^{M^{J\neq0}_{max}}_{eq}$ (km) & $M^{J\neq0}_{max}/M_\odot$ &  $P^{M^{J\neq0}_{max}}$ (sec)\\
\colrule
$\mu=2$  & Chandrasekhar    & $1.07\times 10^{10}$				 &1198.91					& 1583.47				& 1.5159										 & 0.884\\
\hline
    {}   & Salpeter   & $1.07\times 10^{10}$         &1193.08         & 1575.94       & 1.4996                     & 0.883\\
$^{4}$He & RFMT       & $5.46\times 10^{9}$          &1458.58         & 1932.59       & 1.5001                     & 1.199\\
\hline
    {}   & Salpeter   & $1.08\times 10^{10}$         &1183.99         & 1564.16       & 1.4833                     & 0.878\\
$^{12}$C & RFMT       & $6.95\times 10^{9}$          &1349.15         & 1785.98       & 1.4736                     & 1.074\\
\hline
    {}   & Salpeter   & $1.09\times 10^{10}$         &1178.88         & 1556.68       & 1.4773                     & 0.875\\
$^{16}$O & RFMT       & $7.68\times 10^{9}$          &1308.09         & 1730.65       & 1.4667                     & 1.027\\
\hline
    {}   & Salpeter   & $1.14\times 10^9$            &2002.43         & 2693.17       & 1.2050                     & 2.202\\
$^{56}$Fe& RFMT       & $1.18\times 10^9$            &2000.11         & 2686.06       & 1.2017                     & 2.195 \\
\botrule
\end{tabular} 
\caption{The maximum rotating mass of general relativistic uniformly rotating $^4$He, $^{12}$C, $^{16}$O and $^{56}$Fe WDs for different EoS. $\rho_{M^{J\neq0}_{max}}$, $R^{M^{J\neq0}_{max}}_{p}$, $R^{M^{J\neq0}_{max}}_{eq}$, and $P^{M^{J\neq0}_{max}}$ are central density, polar and equatorial radii, and rotation period of the configuration with the maximum mass, ${M^{J\neq0}_{max}}$.}
\label{tab:mrotmax}}
\end{table*}

It is worth mentioning that the maximum mass of RWDs is not associated with a critical maximum density for gravitational collapse. This is in contrast with the non-rotating case where the configuration of maximum mass (turning-point) corresponds to a critical maximum density over which the WD is unstable against gravitational collapse. 

The angular momentum $J$ along the mass-shedding sequence is not constant and thus the turning-point criterion (\ref{eq:Jcons}) does not apply to this sequence. Therefore the configuration of maximum rotating mass (\ref{eq:rotMmax}) does not separate stable from secular axisymmetrically unstable WDs. We have also verified that none of the RWDs belonging to the mass-shedding sequence is a turning-point of some $J=$constant sequence, and therefore they are indeed secularly stable. We therefore extend the Keplerian sequence all the way up to the critical density for inverse $\beta$ decay, $\rho^{\beta}_{\rm crit}$, see Table \ref{tab:betadecay} and Fig.~\ref{fig:MrhoCO}.

\section{The minimum rotation period}\label{sec:6}

The minimum rotation period $P_{min}$ of WDs is obtained for a configuration rotating at Keplerian angular velocity, at the critical inverse $\beta$-decay density; i.e. is the configuration lying at the crossing point between the mass-shedding and inverse $\beta$-decay boundaries, see Figs.~\ref{fig:MrhoCO} and \ref{fig:JMCO}. For $^4$He, $^{12}$C, $^{16}$O, and $^{56}$Fe RWDs we found the minimum rotation periods $\sim 0.28$, $0.50$, $0.69$ and $2.19$ seconds, respectively (see Table \ref{tab:concl} for details). In Table \ref{tab:mrotmax} we compare the properties of the configuration with minimum rotation period using different EOS, namely Chandrasekhar $\mu=2$, Salpeter, and RFMT EOS.

\begin{table*}
\centering
{\begin{tabular}{@{}ccccccc@{}}
Nuclear composition    & EoS        & $\rho^{\beta}_{\rm crit}$ (g/cm$^3$) & $R^{P_{min}}_{p}$ (km) & $R^{P_{min}}_{eq}$ (km) & $M^{J\neq0}_{P_{min}}/M_\odot$ & $P_{min}$ (sec)\\
\colrule
$\mu=2$  & Chandra    & $1.37\times 10^{11}$				 &562.79					& 734.54				& 1.4963										 & 0.281\\
\hline
         & Salpeter   & $1.37\times 10^{11}$         &560.41          & 731.51        & 1.4803                     & 0.281\\
$^{4}$He & RFMT       & $1.39\times 10^{11}$         &563.71          & 735.55        & 1.4623                     & 0.285\\
\hline
         & Salpeter   & $3.88\times 10^{10}$         &815.98          & 1070.87       & 1.4775                     & 0.498\\
$^{12}$C & RFMT       & $3.97\times 10^{10}$         &816.55          & 1071.10       & 1.4618                     & 0.501\\
\hline
         & Salpeter   & $1.89\times 10^{10}$         &1005.62         & 1324.43       & 1.4761                     & 0.686\\
$^{16}$O & RFMT       & $1.94\times 10^{10}$         &1005.03         & 1323.04       & 1.4630                     & 0.687\\
\hline
         & Salpeter   & $1.14\times 10^9$            &2002.43         & 2693.17       & 1.2050                     & 2.202\\
$^{56}$Fe& RFMT       & $1.18\times 10^9$            &2000.11         & 2686.06       & 1.2018                     & 2.195 \\
\botrule
\end{tabular} 
\caption{The minimum rotation period of general relativistic rotating $^4$He, $^{12}$C, $^{16}$O and $^{56}$Fe WDs. $\rho^{\beta}_{\rm crit}$ is the critical density for inverse $\beta$ decay. $M^{J\neq0}_{P_{min}}$, $R^{P_{min}}_{p}$, and $R^{P_{min}}_{e}$ are the mass, polar, and equatorial radii corresponding to the configuration with minimum rotation period, $P_{min}$.}
\label{tab:pmin}}
\end{table*}

In the case of $^{12}$C WDs, the minimum period $0.50$ seconds have to be compared with the value obtained assuming as critical density the threshold for pycnonuclear reactions. Assuming lifetimes $\tau^{\rm C+C}_{pyc}=10$ Gyr and 0.1 Myr, corresponding to critical densities $\rho_{pyc}\sim 9.26\times 10^9$ g cm$^{-3}$ and $\rho_{pyc}\sim 1.59\times 10^{10}$ g cm$^{-3}$, we obtain minimum periods $P^{pyc}_{min}=0.95$ and 0.75 seconds, respectively.

It is interesting to compare and contrast some classical results with the ones presented in this work. Using post-Newtonian approximation, \cite{1966ZA.....64..504R} analyzed the problem of dynamical stability of maximally rotating RWDs, i.e. WDs rotating at the mass-shedding limit. The result was a minimum polar radius of 363 km, assuming the Chandrasekhar EOS  with $\mu=2$. The Roxburgh critical radius is rather small with respect to our minimum polar radii, see Table \ref{tab:concl}. It is clear that such a small radius would lead to a configuration with the central density over the limit established by inverse $\beta$-decay: the average density obtained for the Roxburgh's critical configuration is $\sim 1.47\times 10^{10}$ g/cm$^3$, assuming the maximum mass $1.48 M_\odot$ obtained in the same work (see Table \ref{tab:mrototh} in App.~\ref{app:3}). A configuration with this mean density will certainly have a central density larger than the inverse $\beta$-decay density of $^{12}$C and $^{16}$O, $3.97\times 10^{10}$ g/cm$^3$ and $1.94\times 10^{10}$ g/cm$^3$, respectively (see Table \ref{tab:betadecay}). The rotation period of the WD at the point of dynamical instability of Roxburgh must be certainly shorter than the minimum values presented here. 

The above comparison is in line with the fact that we did not find any turning-point that cross the mass-shedding sequence (see Figs.~\ref{fig:MrhoCO}--\ref{fig:MRCO}). Presumably, ignoring the limits imposed by inverse $\beta$-decay and pycnonuclear reactions, the boundary determined by the turning-points could cross at some higher density the Keplerian sequence. Such a configuration should have a central density very similar to the one found by \cite{1966ZA.....64..504R}.

In the work of \cite{1971Ap......7..274A} the problem of the minimum rotation period of a WD was not considered. However, they showed their results for a range of central densities covering the range of interest of our analysis. Thus, we have interpolated their numerical values of the rotation period of WDs in the Keplerian sequence and calculated the precise values at the inverse $\beta$-decay threshold for $^4$He, $^{12}$C, and $^{16}$O that have $\mu=2$ and therefore in principle comparable to the Chandrasekhar EOS results with the same mean molecular weight. We thus obtained minimum periods $\sim 0.31$, $0.55$, $0.77$ seconds, in agreement with our results (see Table \ref{tab:pmin}).

\begin{figure*}
\centering
\includegraphics[width=0.48\hsize,clip]{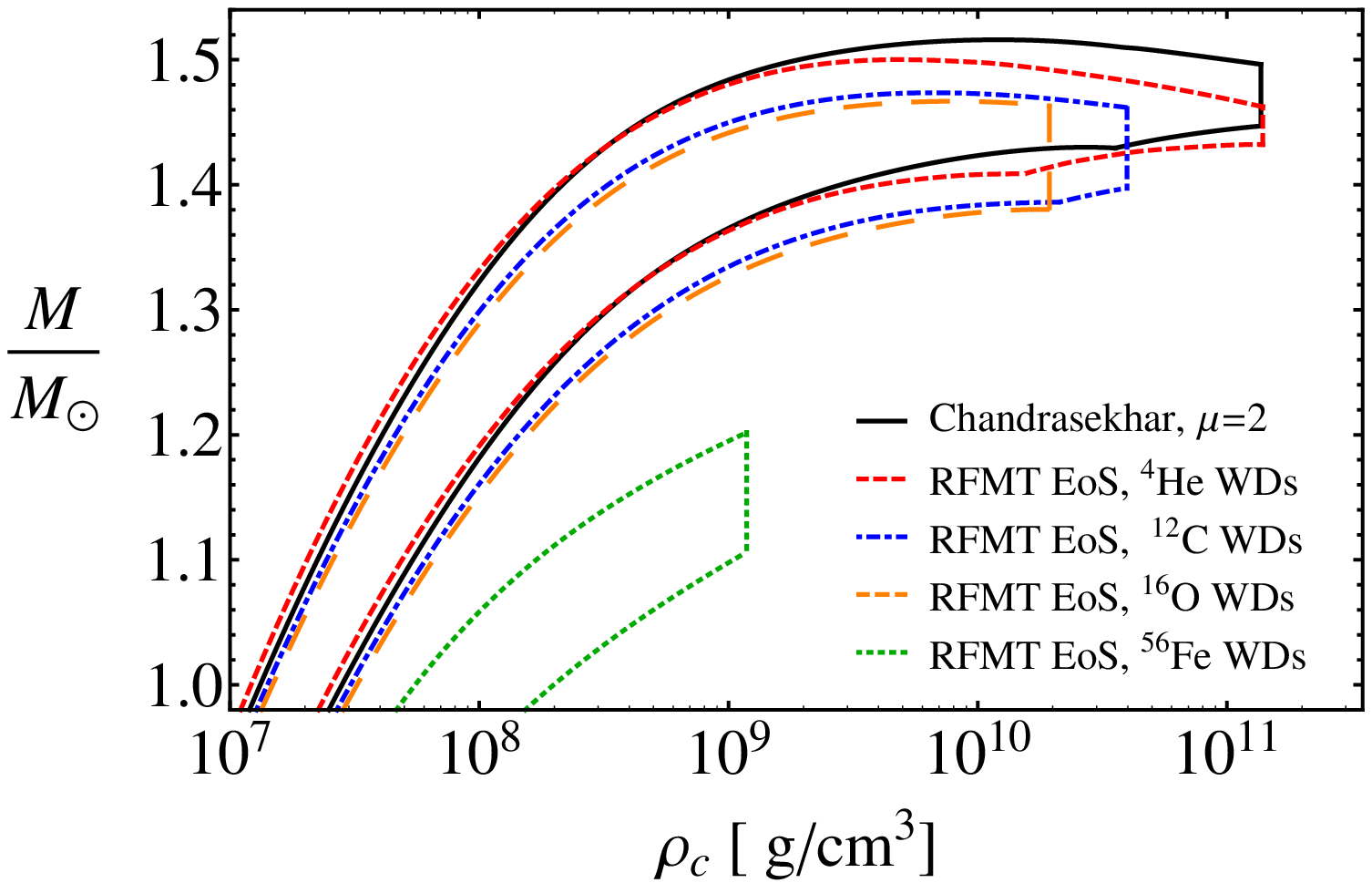} \includegraphics[width=0.48\hsize,clip]{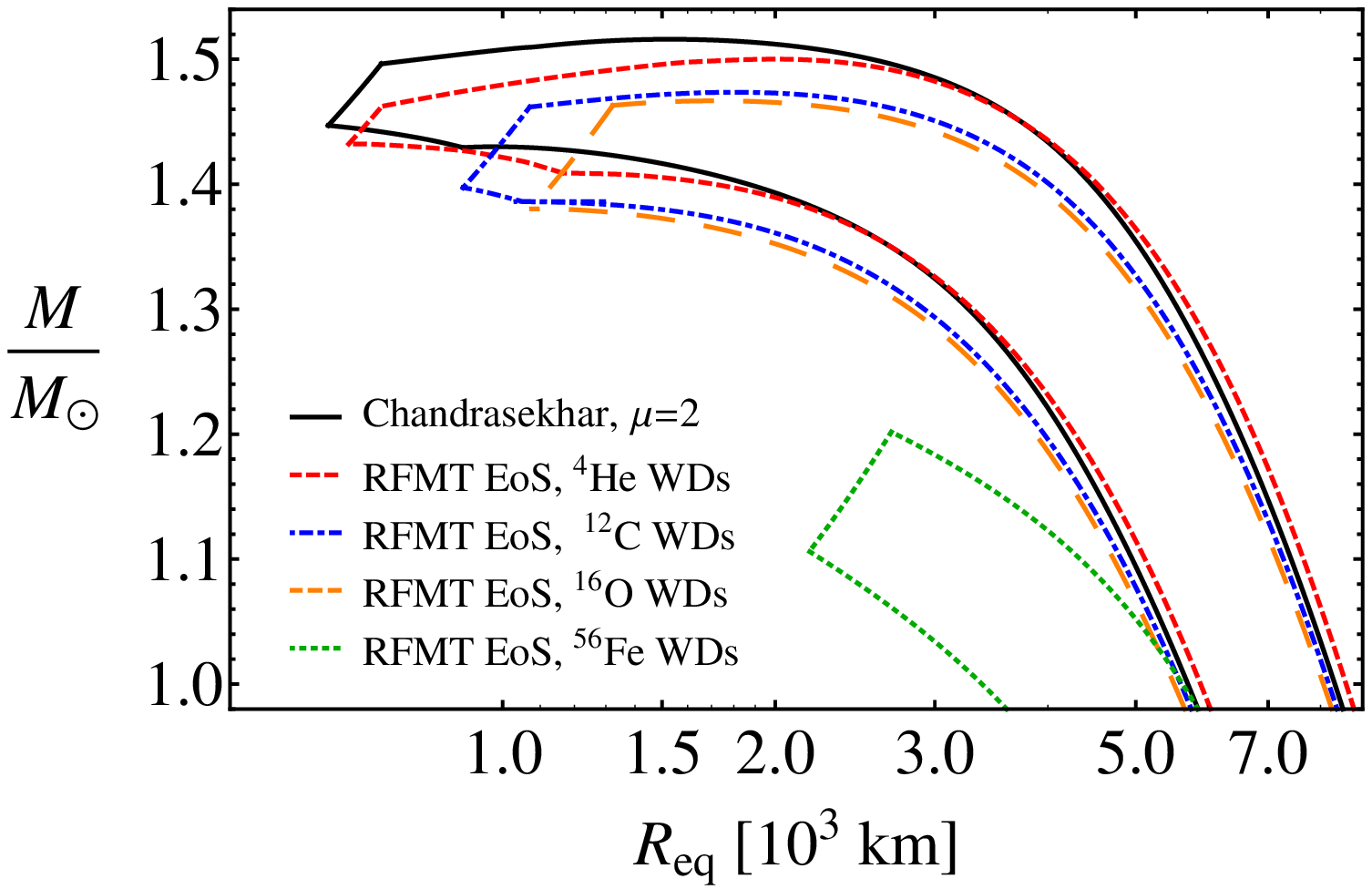}
\caption{Mass versus central density (left panel) and mass versus equatorial radius (right panel) for general relativistic WDs using the Chandrasekhar and the RFMT EOS.}\label{fig:ChandravsRFMT}
\end{figure*}

It is important to stress that, although it is possible to compare the results using the Chandrasekhar EOS $\mu=2$ with the ones obtained for the RFMT EOS, both qualitative and quantitative differences exist between the two treatments. In the former a universal mass-density and mass-radius relation is obtained assuming $\mu=2$ while, in reality, the configurations of equilibrium depend on the specific values of $Z$ and $A$ in non-trivial way. For instance, $^4$He, $^{12}$C, and $^{16}$O have $\mu=2$ but the configurations of equilibrium are rather different. This fact was emphasized by \cite{hamada61} in the Newtonian case and further in GR by \cite{RotD2011}, for non-rotating configurations. In Fig.~\ref{fig:ChandravsRFMT} we present a comparison of the mass-density and mass-radius for the universal Chandrasekhar $\mu=2$ and the RFMT EOS for specific nuclear compositions.

\section{Occurrence of secular axisymmetric instability}\label{sec:7}

Regarding the stability of rotating WDs, \cite{1968ApJ...151.1089O,1969ApJ...155..987O,1975ApJ...199..179D} showed that uniformly rotating Newtonian polytropes and WDs described by the uniform degenerate electron fluid EOS are axisymmetrically stable at any rotation rate. In clear contrast with these results, we have shown here that uniformly RWDs can be indeed be secularly axisymmetric unstable as can be seen from Figs.~\ref{fig:MrhoCO}--\ref{fig:JMCO} (green boundary). We have constructed in App.~\ref{app:3} Newtonian RWDs for the Chandrasekhar EOS and compare the differences with the general relativistic counterpart. Apart from the quantitative differences for the determination of the mass at high densities, it can be seen from Fig.~\ref{fig:NewtvsGRminper} (left panel) the absence of turning-points in the Newtonian mass-density relation. This can be understood from the fact that the maximum stable mass of non-rotating WDs is, in the Newtonian case, reached formally at infinite central density. We should then expect that turning-points will appear only from a post-Newtonian approximation, where the critical mass is shifted to finite densities \citep[see e.g.][for the calculation of dynamical instability for post-Newtonian RWDs obeying the Chandrasekhar EOS]{1966ZA.....64..504R}.

In this respect the Fig.~\ref{fig:JMCO} is of particular astrophysical relevance. Configurations lying in the filled region are stable against mass-shedding, inverse $\beta$-decay and secular axisymmetric instabilities. RWDs with masses smaller than the maximum non-rotating mass (Sub-Chandrasekhar WDs), i.e. $M^{J\neq 0}<M_{max}^{J=0}$, can have angular momenta ranging from a maximum at the mass-shedding limit all the way down to the non-rotating limit $J=0$. SCWDs, however, are stabilized due to rotation and therefore there exist a minimum angular momentum, $J_{min}>0$, to guarantee their stability. We have shown above that secular axisymmetric instability is relevant for the determination of this minimum angular momentum of SCWDs (see green boundary in Fig.~\ref{fig:JMCO}). It is interesting to note in this respect that from our results it turns out that SCWDs with \emph{light chemical compositions} such as $^{4}$He and $^{12}$C, are unstable against \emph{axisymmetric}, inverse $\beta$-decay and mass-shedding instabilities. On the opposite, in SCWDs with \emph{heavier chemical compositions}, such as $^{16}$O and $^{56}$Fe, the secular axisymmetric instability does not take place; see Fig.~\ref{fig:JMCO}. The existence of the new boundary due to secular axisymmetric instability is a critical issue for the evolution of SCWDs since their lifetime might be reduced depending on their initial mass and angular momentum. 

From the quantitative point of view, we have found that axisymmetric instability sets in for $^{12}$C SCWDs in the range of masses $M^{J=0}_{max}<M\lesssim 1.397 M_\odot$, for some specific range of rotation periods $\gtrsim 1.24$ seconds. We can express the minimum rotation period that a SCWD with a mass $M$ within the above mass range can have through the fitting formula
\begin{equation}\label{eq:axiformula}
P_{axi}=0.062 \left(\frac{M-M^{J=0}_{max}}{M_\odot}\right)^{-0.67}\,\,{\rm seconds}\, ,
\end{equation}
where $M^{J=0}_{max}$ is the maximum mass of general relativistic non-rotating $^{12}$C WDs, $M^{J=0}_{max}\approx 1.386 M_\odot$ (see Table \ref{tab:mcrit} and \cite{RotD2011}). Thus, Eq.~(\ref{eq:axiformula}) describes the rotation periods of the configurations along the green-dotted boundary in Figs.~\ref{fig:MrhoCO}, \ref{fig:MRCO}, and \ref{fig:JMCO}. Correspondingly, the central density along this instability boundary varies from the critical density of static $^{12}$C WDs, $\rho^{{\rm C}, J=0}_{\rm crit} = 2.12\times 10^{10}$ g cm$^{-3}$ (see Table \ref{tab:mcrit}), up to the inverse $\beta$-decay density, $\rho^{\rm C}_{\beta}=3.97\times 10^{10}$ g cm$^{-3}$ (see Table \ref{tab:betadecay}).

It is important to note that at the lower edge of the density range for axisymmetric instability, $\rho^{{\rm C}, J=0}_{\rm crit}$, the timescale of C+C pycnonuclear reactions are $\tau^{\rm C+C}_{pyc} \approx 339$ yr (see Fig.~\ref{fig:taupyc}). It becomes then of interest to compare this timescale with the corresponding one of the secular axisymmetric instability that sets in at the same density.

The growing time of the secular instability is given by the dissipation time that can be driven either by gravitational radiation or viscosity \citep{chandrasekhar70}. However, gravitational radiation reaction is expected to drive secular instabilities for systems with rotational to gravitational energy ratio $T/|W|\sim 0.14$, the bifurcation point between McClaurin spheroids and Jacobi ellipsoids \citep[see][for details]{chandrasekhar70}. Therefore, we expect gravitational radiation to become important only for differentially rotating WDs, which can attain more mass and more angular momentum \citep{1968ApJ...151.1089O}. In the present case of general relativistic uniformly RWDs, only the viscosity timescale $\tau_v$ is relevant. A rotating star that becomes secularly unstable first evolve with a characteristic time $\tau_v$ and eventually reach a point of dynamical instability, thus collapsing within a time $\tau_{dyn}\approx \Omega^{-1}_K \sim \sqrt{R^3/G M} \lesssim 1$ s, where $R$ is the radius of the star \citep[see e.g.][]{stergioulas}.

The viscosity timescale can be estimated as $\tau_v=R^2\rho/\eta$ \citep[see e.g.][]{lindblom87}, where $\rho$ and $\eta$ are the density and viscosity of the star. The viscosity of a WD assuming degenerate relativistic electrons is given by \citep{1973ApJ...183..205D}
\begin{equation}\label{eq:viscosity1}
\eta_{fluid} =4.74\times 10^{-2} \frac{H_\Gamma(Z)}{Z} \rho^{5/3}\left[ \left( \frac{\rho}{2\times 10^6}\right)^{2/3}+1 \right]^{-1}\, ,
\end{equation}
where $H_\Gamma(Z)$ is a slowly varying dimensionless contstant that depends on the atomic number $Z$ and the Coulomb to thermal energy ratio
\begin{equation}\label{eq:gamma}
\Gamma=\frac{e^2 Z^2}{k_B T}\left(\frac{4 \pi}{3}\frac{\rho}{2 Z Mu}\right)^{1/3}\, ,
\end{equation}
where $k_B$ is the Boltzmann constant and $A\simeq 2 Z$ has been used.

The expression (\ref{eq:viscosity1}) is valid for values of $\Gamma$ smaller than the critical value for crystallization $\Gamma_{cry}$. The critical $\Gamma_{cry}$ is not well constrained but its value should be of the order of $\Gamma_{cry} \sim 100$ \citep[see e.g.][]{1973ApJ...183..205D,shapirobook}. The critical value $\Gamma_{cry}$ defines a crystallization temperature $T_{cry}$ under which the system behaves as a solid. For $\Gamma_{cry} \sim 100$, we have $T_{cry}\approx 8\times 10^7 [\rho/(10^{10}\,{\rm g\,cm}^{-3})]^{1/3}$ K, for $Z=6$. When $\Gamma$ approaches $\Gamma_{cry}$ the viscosity can increase drastically to values close to \citep{vanhornbook,1973ApJ...183..205D}
\begin{equation}\label{eq:viscosity2}
\eta_{cry} =4.0\times 10^{-2} \left(\frac{Z}{7}\right)^{2/3} \rho^{5/6}\exp[0.1(\Gamma-\Gamma_{cry})]\, .
\end{equation}

For instance, we find that at densities $\rho^{{\rm C}, J=0}_{\rm crit}$ and assuming a central temperature $T\gtrsim 0.5 T_{cry}$ with $T_{cry}\approx 10^8$ K, the viscous timescale is in the range $10\lesssim \tau_v \lesssim 1000$ Myr, where the upper limit is obtained using Eq.~(\ref{eq:viscosity1}) and the lower limit with Eq.~(\ref{eq:viscosity2}). These timescales are longer than the pycnonuclear reaction timescale $\tau^{\rm C+C}_{pyc} = 339$ yr at the same density. So, if the pycnonuclear reaction rates are accurate, it would imply that pycnonuclear reactions are more important to restrict the stability of RWDs with respect to the secular instability. However, we have to keep in mind that, as discussed in Sec.~\ref{sec:3d}, the pycnonuclear critical densities are subjected to theoretical and experimental uncertainties, which could in principle shift them to higher values. For instance, a possible shift of the density for pycnonuclear instability with timescales $\tau^{C+C}_{pyc}\sim 1$ Myr to higher values $\rho^{\rm C+C}_{pyc}>\rho^{{\rm C}, J=0}_{\rm crit}$, would suggest an interesting competition between secular and pycnonuclear instability in the density range $\rho^{{\rm C}, J=0}_{\rm crit}<\rho<\rho^{\rm C}_{\beta}$.

\section{Spin-up and spin-down evolution}\label{sec:8}

It is known that at constant rest-mass $M_0$, entropy $S$ and chemical composition $(Z,A)$, the spin evolution of a RWD is given by \citep[see][for details]{shapiro90}
\begin{equation}
\dot{\Omega} = \frac{\dot{E}}{\Omega} \left( \frac{\partial \Omega}{\partial J} \right)_{M_0,S,Z,A}\, ,
\end{equation}
where $\dot{\Omega} \equiv d\Omega/dt$ and $\dot{E} \equiv dE/dt$, with $E$ the energy of the star. 

Thus, if a RWD is loosing energy by some mechanism during its evolution, that is $\dot{E}<0$, the change of the angular velocity $\Omega$ in time depends on the sign of $\partial \Omega/\partial J$; RWDs that evolve along a track with $\partial \Omega/\partial J>0$, will spin-down ($\dot{\Omega}<0$) and the ones following tracks with $\partial \Omega/\partial J<0$ will spin-up ($\dot{\Omega}>0$). 

\begin{figure*}
\includegraphics[width=0.48\hsize,clip]{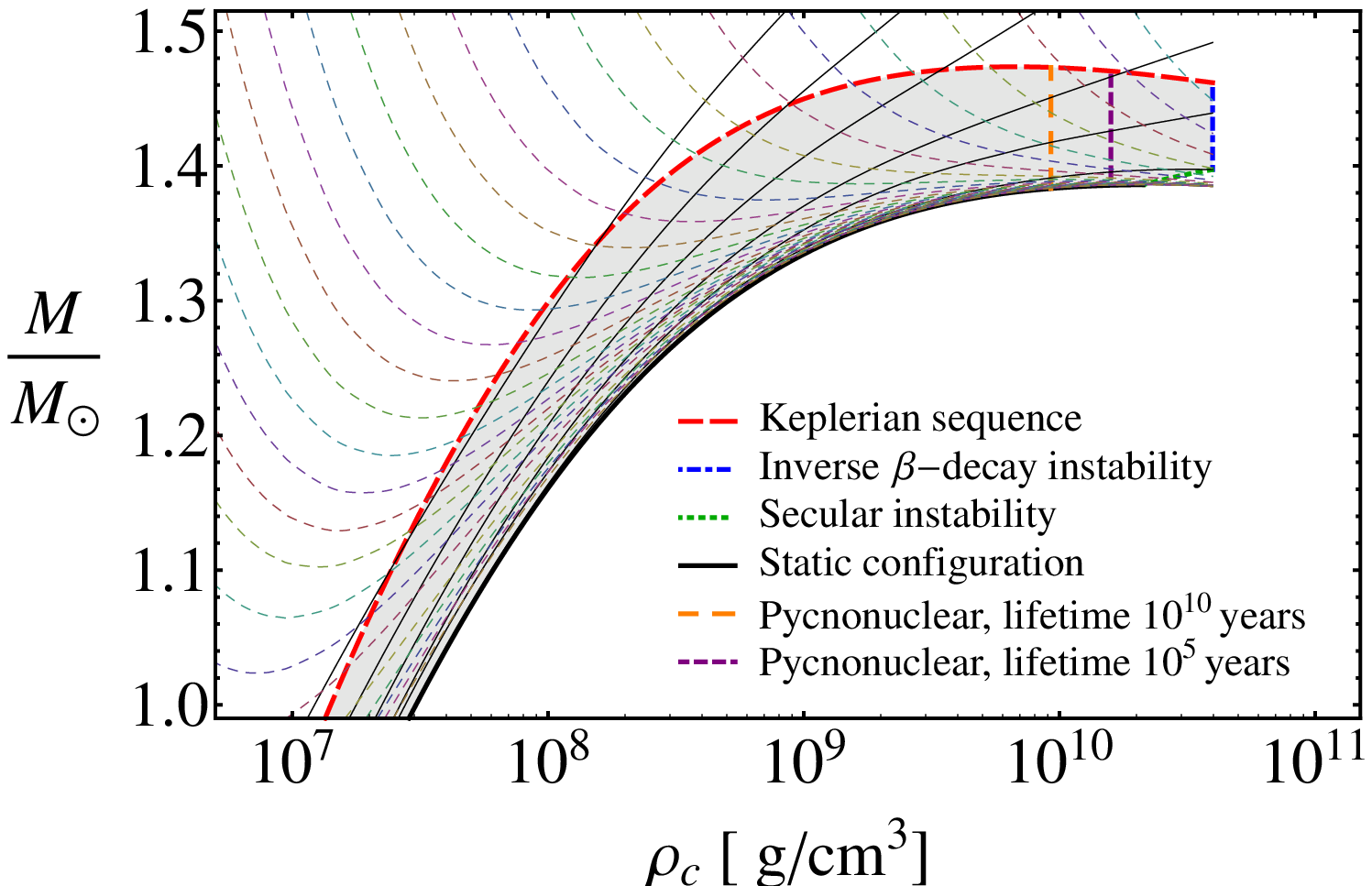} \includegraphics[width=0.48\hsize,clip]{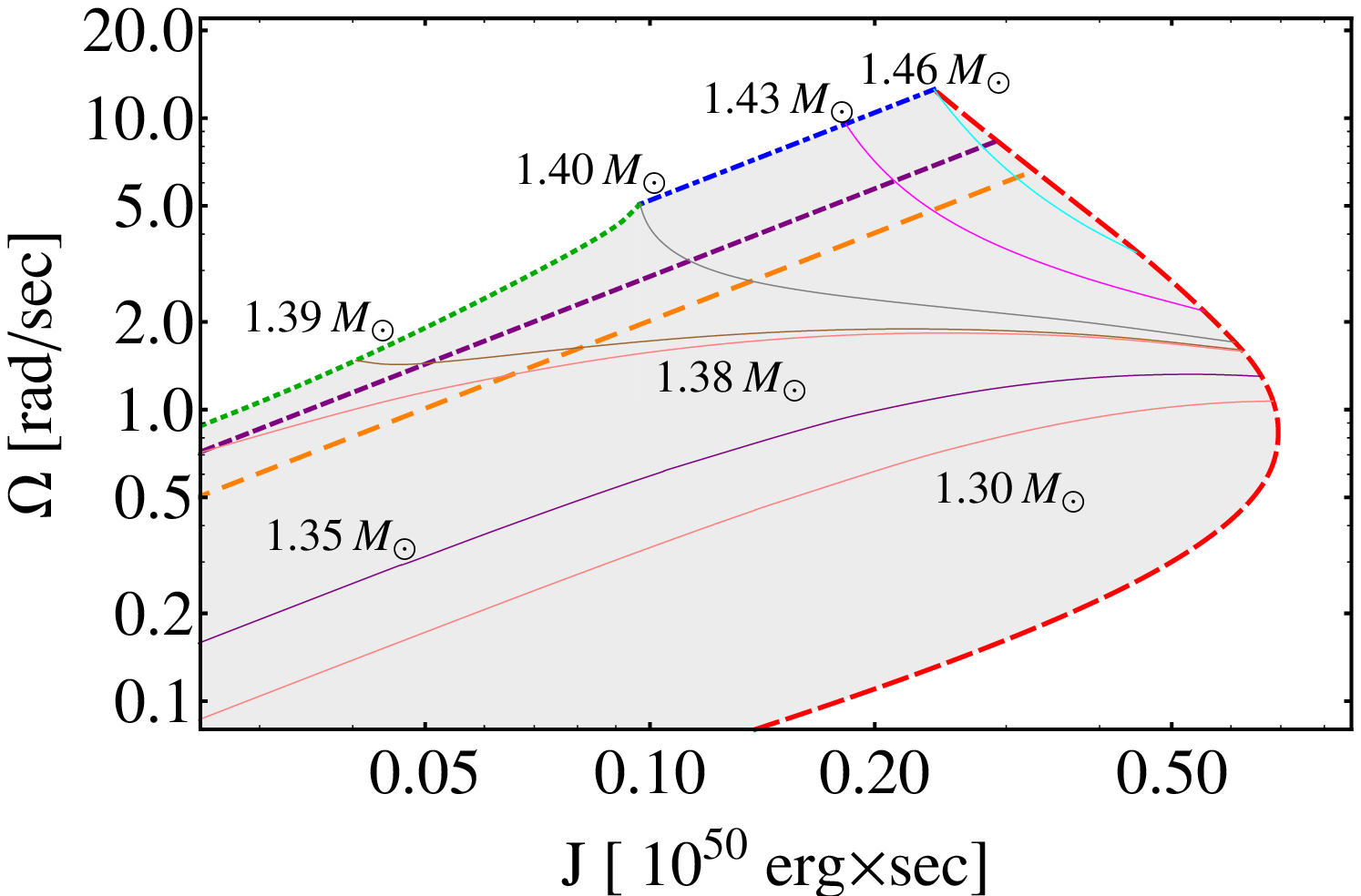}
\caption{Left panel: mass versus the central density for $^{12}$C RWDs. The solid black curves correspond to $J$=constant sequences, where the static case $J=0$ the thickest one. The color thin-dashed curves correspond to $\Omega$=constant sequences. The Keplerian sequence is the red thick dashed curve, the blue thick dotted-dashed curve is the inverse $\beta$-decay instability boundary, and the green thick dotted curve is the axisymmetric secular instability boundary. Right panel: contours of constant rest-mass in the $\Omega-J$ plane; RWDs that evolve along a track with $\partial \Omega/\partial J>0$ spin-down by loosing angular momentum while, the ones with $\partial \Omega/\partial J<0$, spin-up.}\label{fig:consOmJ}
\end{figure*}

In Fig.~\ref{fig:consOmJ} we show, in the left panel, the $\Omega=$constant and $J=$constant sequences in the mass-central density diagram and, in the right panel, contours of constant rest-mass in the $\Omega-J$ plane. 

The sign of $\partial\Omega/\partial J$ can be analyzed from the left panel plot of Fig.~\ref{fig:consOmJ} by joining two consecutive $J=$ constant sequences with an horizontal line and taking into account that $J$ decreases from left to right and from up to down. The angular velocity $\Omega$, instead, decreases from right to left and from up to down for SCWDs and, for sub-Chandrasekhar WDs, from left to right and from up to down. We note that, in the SCWDs region $\Omega=$ constant sequences satisfy $\partial \Omega/\partial \rho_c<0$ while, in the sub-Chandrasekhar region, both $\partial \Omega/\partial \rho_c<0$ and $\partial \Omega/\partial \rho_c>0$ appear (see minima). SCWDs can only either spin-up by angular momentum loss or spin-down by gaining angular momentum. In the latter case, the RWD becomes decompressed with time increasing its radius and moment of inertia, and then SCWDs following this evolution track will end at the mass-shedding limit (see Fig.~\ref{fig:consOmJ}). Some evolutionary tracks of sub-Chandrasekhar WDs and SCWDs are shown in the right panel of Fig.~\ref{fig:consOmJ}. It is appropriate to recall here that \cite{shapiro90} showed that spin-up behavior by angular momentum loss occurs for rapidly rotating Newtonian polytropes if the polytropic index is very close to $n=3$, namely for an adiabatic index $\Gamma \approx 4/3$. It was shown explicitly by \cite{2000ApJ...534..359G} that these conditions are achieved only by Super-Chandrasekhar polytropes.

Besides the confirmation of the above known result for SCWDs in the general relativistic case, we report here the presence of minima $\partial \Omega/\partial \rho_c=0$ for some sub-Chandrasekhar masses (see e.g.~the evolution track of the RWD with $M=1.38 M_\odot$ in the right panel of Fig.~\ref{fig:consOmJ}) which raises the possibility that sub-Chandrasekhar WDs can experience, by angular momentum loss, not only the intuitively spin-down evolution, but also spin-up epochs.

\section{Astrophysical implications}\label{sec:9}

It is appropriate to analyze the astrophysical consequences of the general relativistic RWDs presented in this work.

Most of the observed magnetic WDs are massive; for instance REJ 0317-853 with $M \sim 1.35 M_\odot$ and $B\sim (1.7$--$6.6)\times 10^8$ G \citep[see e.g.][]{1995MNRAS.277..971B,2010A&A...524A..36K}; PG 1658+441 with $M \sim 1.31 M_\odot$ and $B\sim 2.3\times 10^6$ G \citep[see e.g.][]{1983ApJ...264..262L,1992ApJ...394..603S}; and PG 1031+234 with the highest magnetic field $\sim 10^9$ G \citep[see e.g.][]{1986ApJ...309..218S,2009A&A...506.1341K}. However, they are generally found to be slow rotators \citep[see e.g.][]{2000PASP..112..873W}. It is worth mentioning that it has been recently shown by \cite{enrique2012} that such a magnetic WDs can be indeed the result of the merger of double degenerate binaries; the misalignment of the final magnetic dipole moment of the newly born RWD with the rotation axis of the star depends on the difference of the masses of the WD components of the binary. 

The precise computation of the evolution of the rotation period have to account for the actual value at each time of the moment of inertia and the equatorial and polar radii of the WD. Whether magnetic and gravitational radiation braking can explain or not the current relatively long rotation periods of some observed magnetic WDs is an important issue that deserves the appropriate attention and will be addressed elsewhere. 

Magnetic braking of SCWDs has been recently invoked as a possible mechanism to explain the delayed time distribution of type Ia supernovae (SNe) \citep[see][for details]{ilkov2012}: a type Ia SN explosion is delayed for a time typical of the spin-down time scale $\tau_B$ due to magnetic braking, providing the result of the merging process of a WD binary system is a magnetic SCWD rather than a sub-Chandrasekhar one. The characteristic timescale $\tau_B$ of SCWD has been estimated to be $10^7 \lesssim \tau_B \lesssim 10^{10}$ yr for magnetic fields comprised in the range $10^6 \lesssim B \lesssim 10^{8}$ G. A constant moment of inertia $\sim 10^{49}$ g cm$^2$ and a fixed critical(maximum) rotation angular velocity 
\begin{equation}
\Omega_{\rm crit}\sim 0.7 \Omega^{J=0}_{\rm K}=0.7 \sqrt{\frac{G M^{J=0}}{R^3_{M^{J=0}}}}\, ,
\end{equation}
have been adopted \citep{ilkov2012}. 

It is important to recall here that, as discussed in Sec.~\ref{sec:8}, SCWDs spin-up by angular momentum loss, and therefore the reference to a ``spin-down'' time scale for them is just historical. SCWDs then evolve toward the mass-shedding limit, which determines in this case the critical angular velocity for rotational instability. 

If we express $\Omega_{K}^{J\neq0}$ in terms of $\Omega_{K}^{J=0}$ (see App.~\ref{app:1b}), taking into account the values of $j$ and $q$ from the numerical integration, we find for RWDs that the Keplerian angular velocity can be written as
\begin{equation}\label{eq:rangeomega}
\Omega_{K}^{J\neq0}=\sigma\Omega_{K}^{J=0}\, ,
\end{equation}
where the coefficient $\sigma$ varies in the interval [0.78,0.75] in the range of central densities $[10^5,10^{11}]$ g cm$^{-3}$. It is important to mention that the above range of $\sigma$ hold approximately the same independently on the chemical composition of the WD. However, the actual numerical value of the critical angular velocity, $\Omega^{J\neq 0}_{K}$, is different for different compositions owing to the dependence on $(Z,A)$ of mass-radius relation of non-rotating WDs.

Furthermore, as we have shown, the evolution track followed by a SCWD depends strongly on the initial conditions of mass and angular momentum as well as on chemical composition, and evolution of the moment of inertia (see Fig.~\ref{fig:consOmJ} and Sec.~\ref{sec:8} for details). It is clear that the assumption of fixed moment of inertia $I\sim 10^{49}$ g cm$^2$, leads to a spin-down time scale depending only on the magnetic field strength. A detailed computation will lead to a strong dependence on the mass of the SCWD; resulting in a two-parameter family of delayed times $\tau_B(M,B)$. Detailed calculations of the lifetime of SCWDs braking-down due to magnetic dipole radiation are then needed to shed light on this important matter. Theoretical work along these lines is currently in progress and the results will be presented in a forthcoming publication.

Massive fast rotating and highly magnetized WDs have been proposed as an alternative scenario of Soft Gamma Ray Repeaters (SGRs) and Anomalous X-ray Pulsars (AXPs); see \cite{M2012} for details. Within such scenario, the range of minimum rotation periods of massive WDs found in this work, $0.3\lesssim P_{min}\lesssim 2.2$ seconds, depending on the nuclear composition (see Table \ref{tab:pmin}), implies the rotational stability of SGRs and AXPs, which possess observed rotation periods $2 \lesssim P \lesssim 12$ seconds. The relatively long minimum period of $^{56}$Fe RWDs $\sim 2.2$ seconds, implies that RWDs describing SGRs and AXPs have to be composed of nuclear compositions lighter than $^{56}$Fe, e.g.~$^{12}$C or $^{16}$O.

\section{Concluding remarks}\label{sec:10}

We have calculated the properties of uniformly RWDs within the framework of GR using the Hartle formalism and our new EOS for cold WD matter based on the relativistic Feynman-Metropolis-Teller treatment \citep{RotC2011}, which generalizes previous approaches including the EOS of \cite{salpeter61}. A detailed comparison with RWDs described by the Chandrasekhar and the Salpeter EOS has been performed.

We constructed the region of stability of RWDs taking into account the mass-shedding limit, secular axisymmetric instability, inverse $\beta$-decay, and pycnonuclear reaction lifetimes. The latter have been computed using the updated theoretical models of \cite{pycyak2005,pycyak2006}. We found that the minimum rotation periods for $^{4}$He, $^{12}$C, $^{16}$O, and $^{56}$Fe RWDs are $\sim 0.3$, $0.5$, $0.7$ and $2.2$ seconds, respectively (see Table \ref{tab:pmin}). For $^{12}$C WDs, the minimum period $0.5$ seconds needs to be compared with the values $P^{pyc}_{min}=0.75$ and 0.95 seconds, obtained assuming as critical density the threshold for pycnonuclear reactions for lifetimes $\tau^{\rm C+C}_{pyc}=0.1$ Myr and 10 Gyr, respectively. For the same chemical compositions, the maximum masses are $\sim 1.500$, $1.474$, $1.467$, $1.202$ $M_\odot$ (see Table \ref{tab:mrotmax}). These results and additional properties of RWDs can be found in Table \ref{tab:concl}. 

We have presented a new instability boundary of general relativistic SCWDs, over which they become axisymmetrically unstable. We have expressed the range of masses and rotation periods where this occurs through a fitting formula given by Eq.~(\ref{eq:axiformula}). A comparison with Newtonian RWDs in App.~\ref{app:3} show to the conclusion that this new boundary of instability for uniformly rotating WDs is a general relativistic effect.

We showed that, by loosing angular momentum, sub-Chandrasekhar RWDs can experience both spin-up and spin-down epochs while, SCWDs, can only spin-up. These results are particularly important for the evolution of WDs whose masses approach, either from above or from below, the maximum non-rotating mass. The knowledge of the actual values of the mass, radii, and moment of inertia of massive RWDs are relevant for the computation of delay collapse times in the models of type Ia SN explosions. A careful analysis of all the possible instability boundaries as the one presented here have to be taken into account during the evolution of the WD at pre-SN stages.

We have indicated specific astrophysical systems where the results of this work are relevant; for instance the long rotation periods of observed massive magnetic WDs; the delayed collapse of SCWDs as progenitors of type Ia SNe; and the alternative scenario for SGRs and AXPs based on massive RWDs.

We would like to thank the anonymous referees for the many comments and suggestions that improved the presentation of our results. J.A.R. is grateful to Enrique Garc\'ia-Berro and Noam Soker for helpful discussions and remarks on the properties of magnetic WDs resulting from WD mergers and on the relevance of this work for the delayed collapse of Super-Chandrasekhar WDs.



\appendix

\section{The Hartle-Thorne solution and equatorial circular orbits}\label{app:1}

\subsection{The Hartle-Thorne vacuum solution}\label{app:1a}

The HT metric given by Eq.~(\ref{eq:h1}) can be written in an analytic closed-form in the exterior vacuum case in terms of the total mass $M$, angular momentum $J$, and quadrupole moment $Q$ of the rotating star. The angular velocity of local inertial frames $\omega(r)$, proportional to $\Omega$, and the functions $h_0$, $h_2$, $m_0$, $m_2$, $k_2$, proportional to $\Omega^2$, are derived from the Einstein equations \citep[see][for details]{H1967,HT1968}. Thus, the metric can be then written as
\begin{equation}\label{ht1}
\begin{split}
ds^2=\left(1-\frac{2{ M }}{r}\right)\left[1+2k_1P_2(\cos\theta)+2\left(1-\frac{2{ M}}{r}\right)^{-1}\frac{J^{2}}{r^{4}}(2\cos^2\theta-1)\right]dt^2+\frac{4J}{r}\sin^2\theta dt d\phi\qquad\qquad\qquad
\\\qquad\qquad\qquad-\left(1-\frac{2{ M}}{r}\right)^{-1}\left[1-2\left(k_1-\frac{6 J^{2}}{r^4}\right)P_2(\cos\theta) -2\left(1-\frac{2{ M}}{r}\right)^{-1}\frac{J^{2}}{r^4}\right]dr^2-r^2[1-2k_2P_2(\cos\theta)](d\theta^2+\sin^2\theta d\phi^2)\,
\end{split}
\end{equation}
where
\begin{equation}\label{ht2}
k_1=\frac{J^{2}}{{ M}r^3}\left(1+\frac{{ M}}{r}\right)+\frac{5}{8}\frac{Q-J^{2}/{ M}}{{ M}^3}Q_2^2\left(\frac{r}{{ M}}-1\right)\ ,\quad 
k_2=k_1+\frac{J^{2}}{r^4}+\frac{5}{4}\frac{Q-J^{2}/{ M}}{{ M}^2r}\left(1-\frac{2{ M}}{r}\right)^{-1/2}Q_2^1\left(\frac{r}{ M}-1\right)\ ,\nonumber
\end{equation}
and
\begin{equation}\label{legfunc}
Q_{2}^{1}(x)=(x^{2}-1)^{1/2}\left[\frac{3x}{2}\ln\frac{x+1}{x-1}-\frac{3x^{2}-2}{x^{2}-1}\right],
\ \ Q_{2}^{2}(x)=(x^{2}-1)\left[\frac{3}{2}\ln\frac{x+1}{x-1}-\frac{3x^{3}-5x}{(x^{2}-1)^2}\right],
\end{equation}
are the associated Legendre functions of the second kind, with $x=r/M -1$, and $P_2(\cos\theta)=(1/2)(3\cos^2\theta-1)$ is the Legendre polynomial. The constants $M$, $J$ and $Q$ the total mass, angular momentum and mass quadrupole moment of the rotating object, respectively. This form of the metric corrects some misprints of the original paper by \cite{HT1968} (see also \cite{berti2005} and \cite{bini2009}). The precise numerical values of $M$, $J$ and $Q$ are calcualted from the matching procedure of the exterior and interior metrics at the surface of the star. 

The total mass of a rotating configuration is defined as $M=M^{J\neq0}=M^{J=0}+\delta M$, where $M^{J=0}$ is the mass of non-rotating configuration and $\delta M$ is the change in mass of the rotating from the non-rotating configuration with the same central density. It should be stressed that in the terms involving $J^2$ and $Q$ the total mass $M$ can be substituted by $M^{J=0}$ since $\delta M$ is already a second order term in the angular velocity.

\subsection{Angular velocity of equatorial circular orbits}\label{app:1b}

The four-velocity $u$ of a test particle on a circular orbit  in equatorial plane of axisymmetric stationary spacetime can be parametrized by the constant angular velocity $\Omega$ with respect to an observer at infinity
\begin{equation}
u=\Gamma[\partial_t+\Omega\partial_{\phi}],
\end{equation}
where $\Gamma$ is a normalization factor which assures that $u^{\alpha}u_{\alpha}=1$. From normalization and geodesics conditions we obtain the following expressions for $\Gamma$ and $\Omega=u^{\phi}/u^{t}$
\begin{eqnarray}\label{eight}
\Gamma&=&\pm(g_{tt}+2\Omega g_{t\phi}+\Omega^2 g_{\phi\phi})^{-1/2},\quad g_{tt,r}+2\Omega g_{t\phi,r}+\Omega^2 g_{\phi\phi,r}=0,
\end{eqnarray}
hence, $\Omega$, the solution of (\ref{eight})$_2$, is given by
\begin{equation}
\Omega_{\pm orb}(r)=\frac{u^{\phi}}{u^{t}}=\frac{-g_{t\phi,r}\pm\sqrt{(g_{t\phi,r})^2-g_{tt,r}g_{\phi\phi,r}}}{g_{\phi\phi,r}},
\end{equation}
where $(+/-)$ stands for co-rotating/counter-rotating orbits, $u^{\phi}$ and $u^{t}$ are the angular and time components of the four-velocity, and a colon stands for partial derivative with respect to the corresponding coordinate. In our case one needs to consider only co-rotating orbits (omitting the plus sign in $\Omega_{+ orb}(r)=\Omega_{orb}(r)$) to determine the mass shedding (Keplerian) angular velocity on the surface of the WD. For the Hartle-Thorne external solution Eq.~(\ref{ht1}) we have
\begin{equation}\label{eq:omegaorb}
\Omega_{orb}(r)=\sqrt{\frac{M}{r^3}}\left[1- j F_{1}(r)+j^2F_{2}(r)+q F_{3}(r)\right],
\end{equation}
where $j=J/M^2$ and $q=Q/M^3$ are the dimensionless angular momentum and quadrupole moment,
\begin{eqnarray*}
&&F_{1}=\left(\frac{M}{r}\right)^{3/2},\quad F_{2}=\frac{48{M}^7-80{M}^6r+4{M}^5r^2-18{M}^4r^3+40{M}^3r^4+10{M}^2r^5+15{M}r^6-15r^7}{16{M}^2r^4(r-2{M})}+F, \\
&&F_{3}=\frac{6{M}^4-8{M}^3r-2{M}^2r^2-3{M}r^3+3r^4}{16{M}^2r(r-2{M})/5}-F,\qquad F=\frac{15(r^3-2{M^3})}{32{M}^3}\ln\frac{r}{r-2{M}}.
\end{eqnarray*}

The mass shedding limiting angular velocity of a rotating star is the Keplerian angular velocity evaluated at the equator ($r=R_{eq}$), i.e.
\begin{equation}\label{eq:omegaK1}
\Omega_K^{J\neq0}=\Omega_{orb}(r=R_{eq}).
\end{equation}

In the static case i.e. when $j=0$ hence $q=0$ and $\delta M=0$ we have the well-known Schwarzschild solution and the orbital angular velocity for a test particle  $\Omega_{ms}^{J=0}$ on the surface ($r=R$) of the WD is given by
\begin{equation}\label{eq:omegaK2}
\Omega_K^{J=0}=\sqrt{\frac{M^{J=0}}{R_{M^{J=0}}^3}}.
\end{equation}

\subsection{Weak field limit}\label{app:1c}

Let us estimate the values of $j$ and $q$ recovering physical units with $c$ and $G$. The dimensionless angular momentum is
\begin{equation}
j=\frac{cJ}{GM^2}=\frac{c}{G}\frac{\alpha MR^2 \Omega}{M^2}=\alpha\left(\frac{\Omega R}{c}\right)\left(\frac{GM}{c^2R}\right)^{-1},
\end{equation}
where we have used the fact that $J=I\Omega$, with $I=\alpha MR^2$, and $\alpha\sim 0.1$ from our numerical integrations. For massive and fast rotating WDs we have $(\Omega R)/c\sim10^{-2}$ and $(GM)/(c^2R)\sim10^{-3}$, so $j\sim 1$.

The dimensionless quadrupole moment $q$ is
\begin{equation}
q=\frac{c^4}{G^2}\frac{Q}{M^3}=\frac{c^4}{G^2}\frac{\beta MR^2}{M^3}=\beta\left(\frac{GM}{c^2R}\right)^{-2},
\end{equation}
where we have expressed the mass quadrupole moment $Q$ in terms of mass and radius of the WD, $Q=\beta MR^2$, where $\beta\sim 10^{-2}$, so we have $q\sim10^4$.

The large values of $j$ and $q$ might arise some suspicion on the products $j F_1$, $j^2F_2$ and $qF_3$ as real correction factors in Eq.~(\ref{eq:omegaorb}). It is easy to check this in the weak field limit $M/r\ll1$, where the functions $F_i$ can be expanded as a power-series
\begin{eqnarray*}
F_{1}=\left(\frac{M}{r}\right)^{3/2},\quad F_{2}\approx\frac{1}{2}\left(\frac{M}{r}\right)^3-\frac{117}{28}\left(\frac{M}{r}\right)^4-6\left(\frac{M}{r}\right)^5-..., \quad F_{3}\approx\frac{3}{4}\left(\frac{M}{r}\right)^2+\frac{5}{4}\left(\frac{M}{r}\right)^3+\frac{75}{28}\left(\frac{M}{r}\right)^4+6\left(\frac{M}{r}\right)^5+...
\end{eqnarray*}
so evaluating at $r=R$
\begin{equation}
j F_1=\alpha\left(\frac{\Omega R}{c}\right)\left(\frac{GM}{c^2R}\right)^{1/2}, \quad j^2F_2=\frac{\alpha}{2}\left(\frac{\Omega R}{c}\right)\left(\frac{GM}{c^2R}\right)^{2}, 
\end{equation}
so we finally have $jF_1\sim10^{-9/2}$, $j^2F_2\sim10^{-9}$, and $qF_3\sim10^{-2}$. We can therefore see that the products are indeed corrections factors and, in addition, that effect due to the quadrupolar deformation is larger than the frame-dragging effect.

\section{Pycnonuclear fusion reaction rates}\label{app:2}

The theoretical framework for the determination of the pycnonuclear reaction rates was developed by \cite{salpeter1969}. The number of reactions per unit volume per unit time can be written as
\begin{equation}
R_{pyc}=Z^4 A\rho S(E_p) 3.90\times 10^{46} \lambda^{7/4} \exp(-2.638/\sqrt{\lambda})\,\,{\rm cm}^{-3}\,{\rm s}^{-1}\, ,\qquad \lambda=\frac{1}{Z^2 A^{4/3}}\left(\frac{\rho}{1.3574\times 10^{11}\,{\rm g}\,{\rm cm}^{-3}}\right)^{1/3}\, ,
\end{equation}
where $S$ are astrophysical factors in units of Mev barns (1 barn=$10^{-24}$ cm$^2$) that have to be evaluated at the energy $E_p$ given by Eq.~(\ref{eq:Ep}). 

For the $S$-factors we adopt the results of \cite{pycyak2005} calculated with the NL2 nuclear model parameterization. For center of mass energies $E\geq 19.8$ MeV, the $S$-factors can be fitted by
\begin{equation}\label{eq:sfactor}
S(E)=5.15\times 10^{16}\exp\left[-0.428 E-\frac{3 E^0.308}{1+e^{0.613 (8-E)}}\right]\,\,{\rm MeV\,barn}\, ,
\end{equation}
which is appropriate for the ranges of the zero-point energies at high densities. For instance, $^{12}$C nuclei at $\rho=10^{10}$ g cm$^{-3}$ have a zero-point oscillation energy $E_p\sim 34$ keV.

All the nuclei $(Z,A)$ at a given density $\rho$ will fuse in a time $\tau_{pyc}$ given by
\begin{equation}\label{eq:taupyc}
\tau_{pyc}=\frac{n_N}{R_{pyc}}=\frac{\rho}{A M_u R_{pyc}}\, ,
\end{equation}
where $n_N=\rho/(A M_u)$ is the ion-density. \cite{pycyak2005} estimated that the $S$-factors (\ref{eq:sfactor}) are uncertain within a factor $\sim 3.5$; it is clear from the above equation that for a given lifetime $\tau_{pyc}$ such uncertainties reflect also in the determination of the density threshold.


\section{Comparison with the Newtonian treatment and other works}\label{app:3}

\begin{figure}
\centering
\includegraphics[width=0.48\hsize,clip]{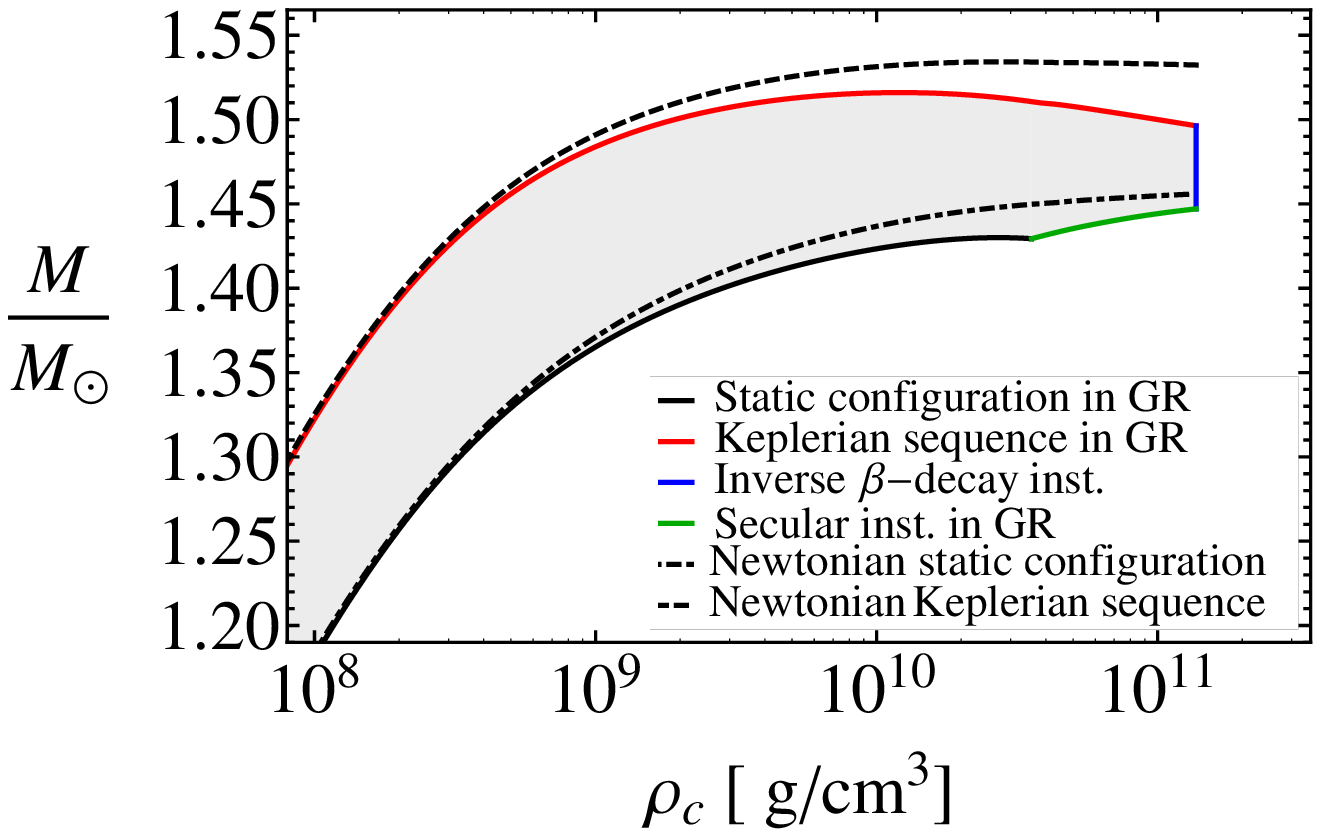} \includegraphics[width=0.48\hsize,clip]{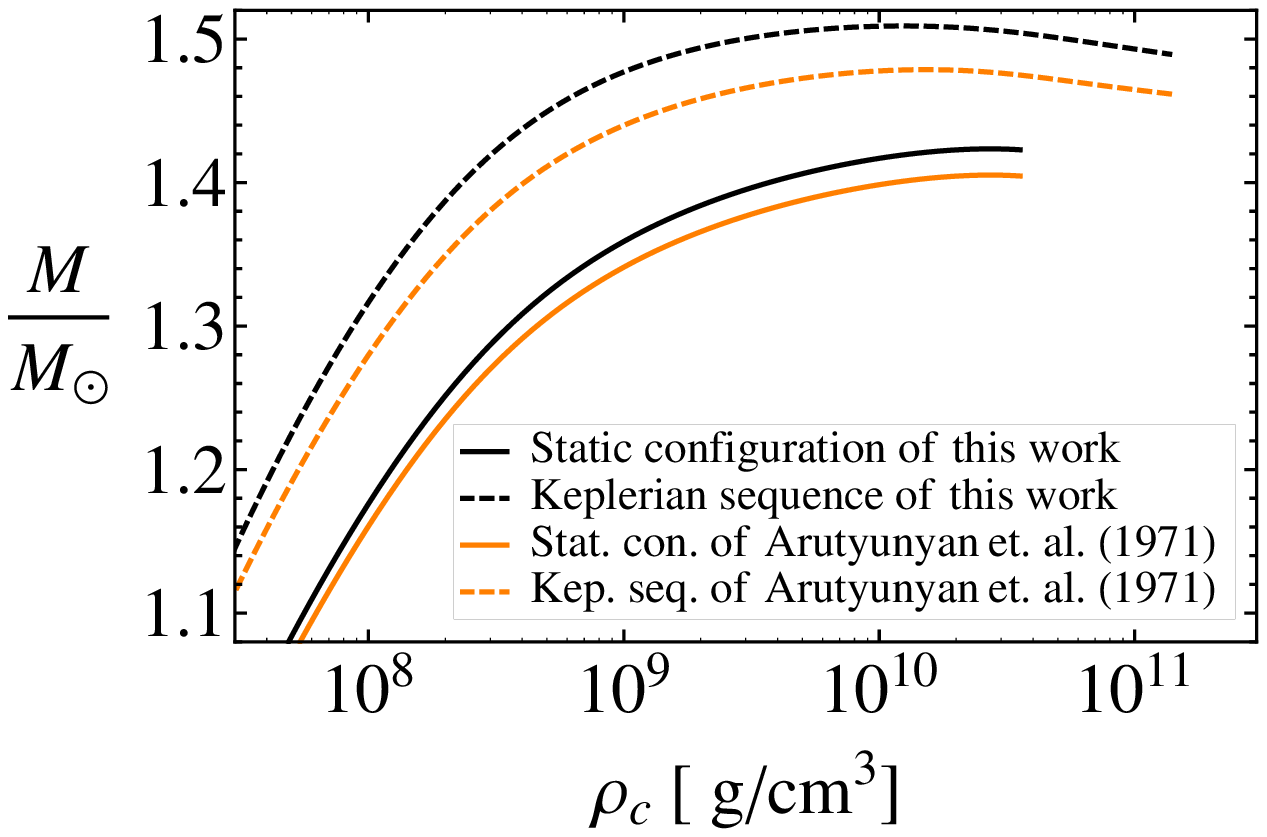}
\caption{Left panel: Mass versus central density of Newtonian and general relativistic WDs for the Chandrasekhar EOS with $\mu=2$. Both the non-rotating case and the Keplerian sequence are shown. We have stopped the density, just for sake of comparison, at the critical density for the onset of inverse $\beta$-decay of $^4$He $\rho=1.39\times 10^{11}$ g cm$^{-3}$. Right panel: Mass versus central density relation for general relativistic WDs for the Chandrasekhar EOS with $\mu=2$ for the static and the Keplerian sequence in this work and the one of \cite{1971Ap......7..274A}.}\label{fig:NewtvsGRminper}
\end{figure}

We have constructed solutions of the Newtonian equilibrium equations for RWDs accurate up to order $\Omega^2$, following the procedure of \cite{H1967}. In Fig.~\ref{fig:NewtvsGRminper} (left panel) we compare these Newtonian configurations with general relativistic RWDs for the Chandrasekhar EOS with $\mu=2$. We can see clearly the differences between the two mass-density relations toward the high density region, as expected. A most remarkable difference is the existence of axisymmetric instability boundary in the general relativistic case, absent in its Newtonian counterpart.

Up to our knowledge, the only previous work on RWDs within GR is the one of \cite{1971Ap......7..274A}. A method to compute RWDs configurations accurate up to second order in $\Omega$ was developed by two of the authors \citep[see][for details]{sedrakyan1968}, independently of the work of \cite{H1967}. In \citep{1971Ap......7..274A}, RWDs were computed for the Chandrasekhar EOS with $\mu=2$. 

In Fig.~\ref{fig:NewtvsGRminper} (right panel) we show the mass-central density relation obtained with their method with the ones constructed in this work for the same EOS. We note here that the results are different even at the level of static configurations, and since the methods are based on construction of rotating configurations from seed static ones, those differences extrapolate to the corresponding rotating objects. This fact is to be added to the possible additional difference arising from the different way of approaching the order $\Omega^2$ in the approximation scheme. The differences between the two equilibrium configurations are evident.

Turning now to the problem of the maximum mass of a RWD, in Table \ref{tab:mrototh} we present the previous results obtained in Newtonian, Post-Newtonian approach and GR by several authors. Depending on their method, approach, treatment, theory and numerical code the authors showed different results. These maximum mass of RWDs are to be compared with the ones found in this work and presented in Table \ref{tab:mrotmax} for the Chandrasekhar $\mu=2$, Salpeter, and RFMT EOS.
\begin{table}
\centering
{
\begin{tabular}{c  c c c}
%
Treatment/EOS & $M^{J\neq0}_{max}/M_{\odot}$ & References \\
\colrule
Newtonian/Chandrasekhar $\mu=2$          &  1.474      & \cite{1965PNAS...54...23A}             \\ 
Newtonian/Polytrope $n=3$         &  1.487      & \cite{1965ZA.....62..134R}          \\ 
Post-Newtonian/Chandrasekhar $\mu=2$     & 1.482      & \cite{1966ZA.....64..504R}  \\ 
GR/Chandrasekhar $\mu=2$                & 1.478      & \cite{1971Ap......7..274A}  \\ 
\botrule
\end{tabular}\label{tab:mrototh}}
\caption{Maximum rotating mass of WDs in literature.}
\end{table}

%

\section{Accuracy of the Hartle's approach}\label{app:4}

In his classic work, \cite{H1967} described the slow rotation regime by requesting that fractional changes in pressure, energy density, and gravitational field due to the rotation of the star are all much smaller with respect to a non-rotating star with the same central density. From a dimensional analysis, such a condition implies
\begin{equation}\label{eq:d1}
\Omega^2\ll\left(\frac{c}{R}\right)^2 \frac{GM^{J=0}}{c^2R}\, ,
\end{equation}
where $M^{J=0}$ is the mass of the unperturbed configuration and $R$ its radius. The expression on the right is the only multiplicative combination of $M, R, G$, and $c$, and in the Newtonian limit coincides with the critical Keplerian angular velocity $\Omega^{J=0}_{K}$ given by Eq.~(\ref{eq:omegaK2}). For unperturbed configurations with $(GM)/(c^2R)<1$, the condition (\ref{eq:d1}) implies $\Omega R/c\ll1$. Namely, every particle must move at non-relativistic velocities if the perturbation to the original geometry have to be small in terms of percentage. Eq.~(\ref{eq:d1}) can be also written as 
\begin{equation}\label{eq:cond}
\Omega \ll \Omega^{J=0}_{K}\, ,
\end{equation}
which is the reason why it is often believed that the slow rotation approximation is not suitable for the description of stars rotating at their mass-shedding value. 

Let us discuss this point more carefully. It is clear that the request that the contribution of rotation to pressure, energy density, and gravitational field to be small can be summarized in a single expression, Eq.~(\ref{eq:d1}), since all of them are quantitatively given by the ratio between the rotational and the gravitational energy of the star. The rotational energy is $T \sim M R^2 \Omega^2$ and the gravitational energy is $|W| \sim G M^2/R=(G M/c^2 R) M c^2$, hence the condition $T/|W|\ll 1$ leads to Eq.~(\ref{eq:d1}) or (\ref{eq:cond}). Now we will discuss the above condition for realistic values of the rotational and gravitational energy of a rotating star, abandoning the assumption of either fiducial or order of magnitude calculations. We show below that the actual limiting angular velocity on the right-hand-side of the condition (\ref{eq:cond}) has to be higher than the Keplerian value.

We can write the gravitational binding energy of the star as $|W|=\gamma G M^2/R$ and the rotational kinetic energy as $T=(1/2)I\Omega^2=(1/2)\alpha M R^2 \Omega^2$, where the constants $\gamma$ and $\alpha$ are structure constants that depends on the density and pressure distribution inside the star. According to the slow rotation approximation, $T/|W|\ll 1$, namely
\begin{equation}
\frac{T}{|W|}=\frac{\alpha M R^2\Omega^2/2}{\gamma G M^2/R}=\left(\frac{\alpha}{2\gamma}\right)\left(\frac{GM}{R^3}\right)^{-1}\Omega^2=\left(\frac{\alpha}{2\gamma}\right)\left(\frac{\Omega}{\Omega_{K}^{J=0}}\right)^2 \ll1,
\end{equation}
which can be rewritten in analogous form to Eq.~{\ref{eq:cond}} as
\begin{equation}\label{eq:cond2}
\Omega\ll\sqrt{\frac{2\gamma}{\alpha}}\Omega_{K}^{J=0}.
\end{equation}

Now we check that the ratio of the structural constants is larger than unity. Let us first consider the simplest example of a constant density sphere. In this case $\alpha=2/5$ and $\gamma=3/5$, so $\sqrt{2\gamma/\alpha}\approx 1.73$, and the condition (\ref{eq:cond2}) is $\Omega\ll 1.73\Omega_{K}^{J=0}$. If we consider now a more realistic density profile, for instance, a polytrope of index $n=3$, we have \citep[see e.g.][]{shapirobook}
\begin{equation}
|W|=\frac{3}{5-n}\frac{GM^2}{R}=\frac{3}{2}\frac{GM^2}{R},\qquad T=\frac{1}{2}I\Omega^2=\frac{1}{2}\frac{2}{3}M\langle r^2\rangle\Omega^2\, 
\end{equation}
where $\langle r^2\rangle=0.11303 R^2$. Therefore we have in this case $\gamma= 3/2$ and $\alpha=0.075$, and so Eq.~(\ref{eq:cond2}) becomes $\Omega\ll 6.32\Omega_{K}^{J=0}$. This is not surprising since $T/|W|\to 0.025$ when $\Omega\to\Omega^{J=0}_{K}$. 

The above analysis has been done assuming spherical symmetry. When deviations from the spherical shape are taken into account, the ratio $T/|W|$ turn to be even smaller than the previous estimates based on spherical polytropes. Since the equatorial radius satisfies $R_{eq}>R$, at mass-shedding we will have $\Omega<\Omega^{J=0}_{K}$. In fact, in the Roche model the mass-shedding angular velocity is $\Omega^{J\neq 0}_{K}=(2/3)^{3/2}\Omega^{J=0}_{K}\approx 0.544 \Omega^{J=0}_{K}$, corresponding to a rotational to gravitational energy ratio $T/|W|\approx 0.0074$ \citep[see e.g.][]{shapirobook}. 

\begin{figure}
\centering
\includegraphics[width=0.48\hsize,clip]{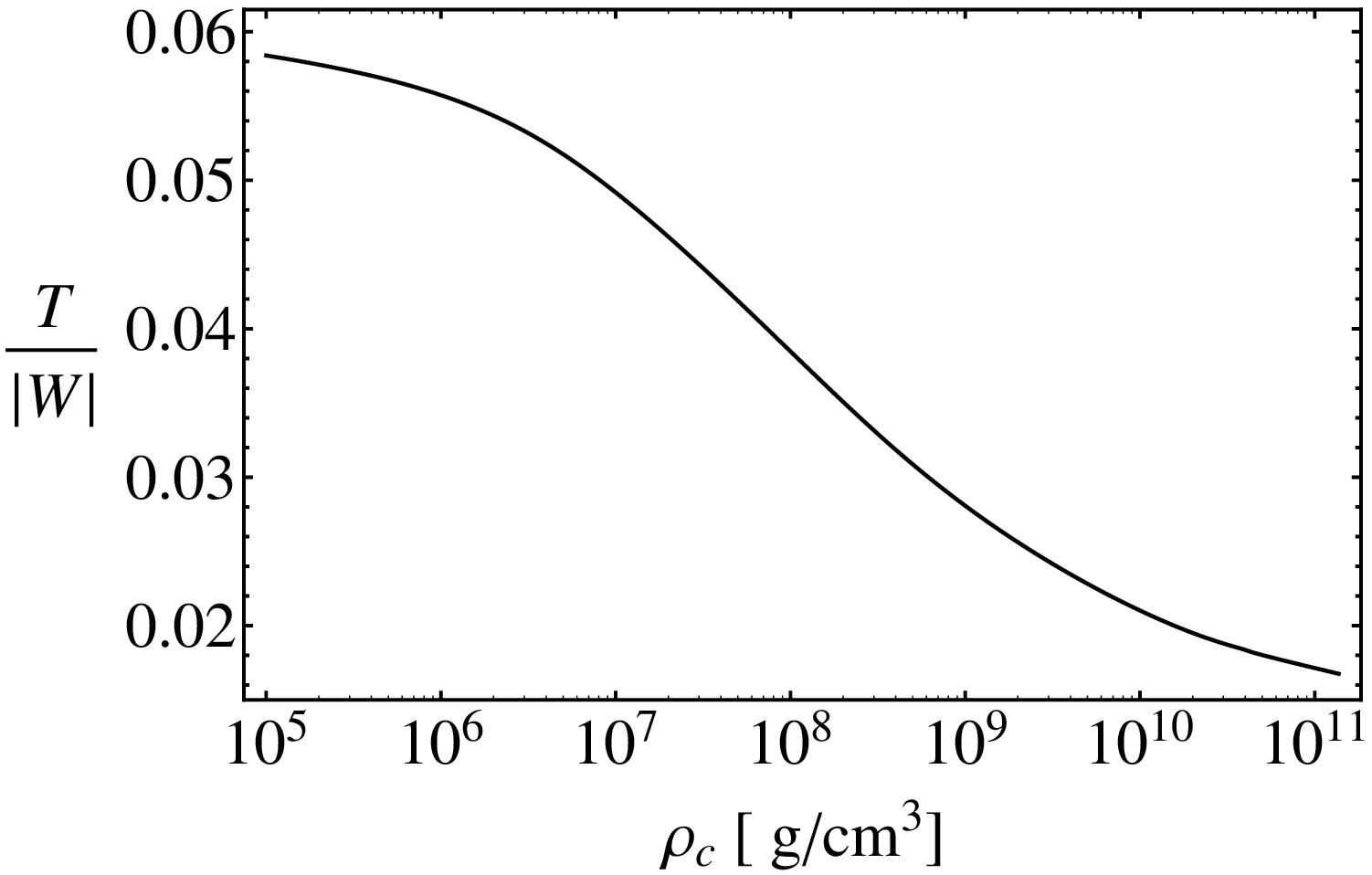} \includegraphics[width=0.48\hsize,clip]{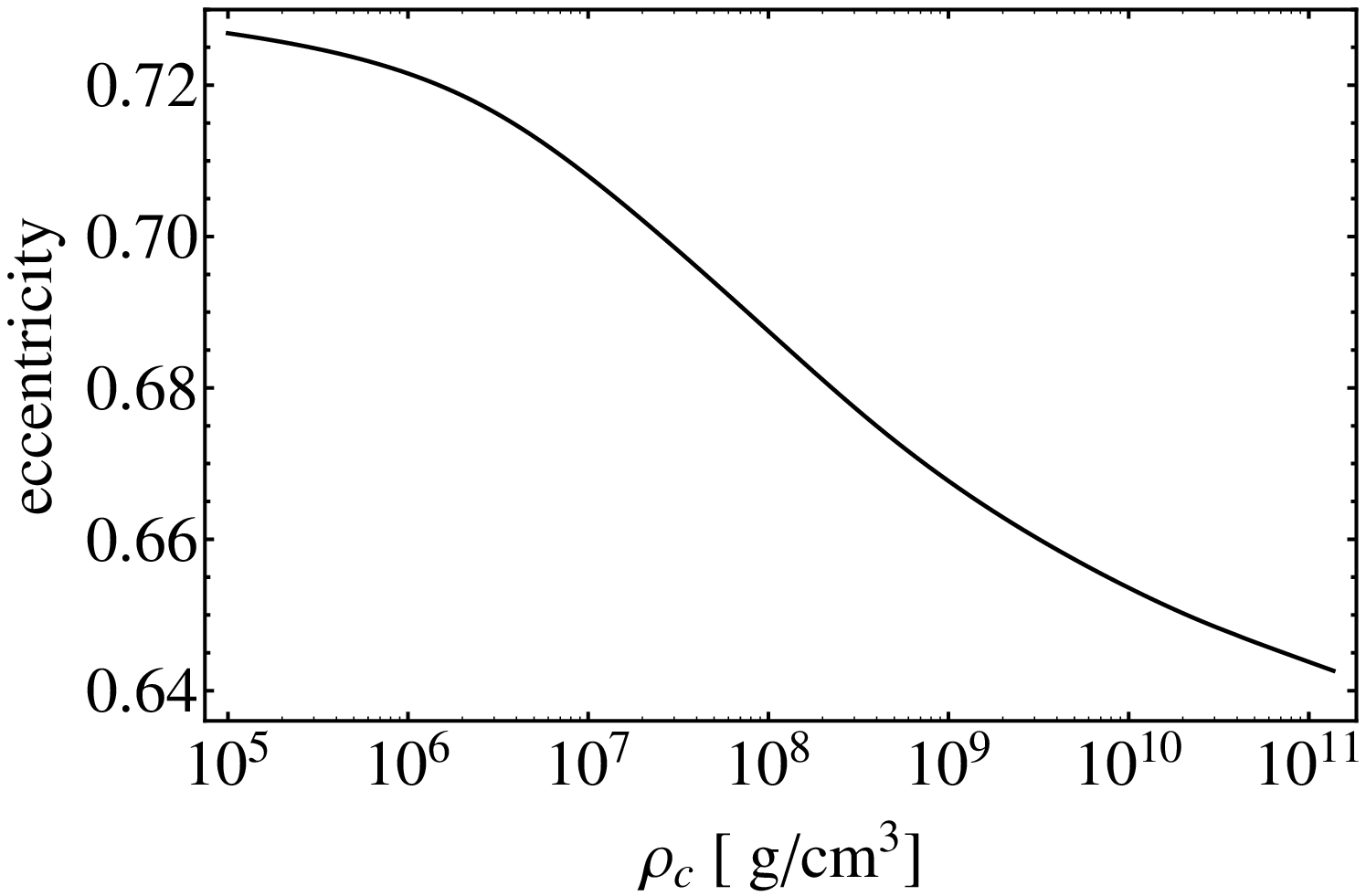}
\caption{Left panel: rotational to gravitational energy ratio versus the central density for maximally rotating RWDs, calculated with the Chandrasekhar EOS $\mu=2$. Right panel: the eccentricity versus the central density for the same sequence of RWDs.}\label{fig:ToverW}
\end{figure}

In our RWDs we have obtained that the mass-shedding angular velocity satisfies $\Omega_{K}^{J\neq0}\approx 0.75\Omega_{K}^{J=0}$ at any density; see Eq.~(\ref{eq:rangeomega}). Accordingly to this, we show in the left panel of Fig.~\ref{fig:ToverW} the ratio $T/|W|$ for RWDs as a function of the central density for the Keplerian sequence. For an increasing central density $T/|W|$ decreases. On the right panel we have plotted the eccentricity versus the central density. For increasing central density the eccentricity decreases, so RWDs become less oblate at higher densities.

\begin{figure}
\centering
\includegraphics[width=0.48\hsize,clip]{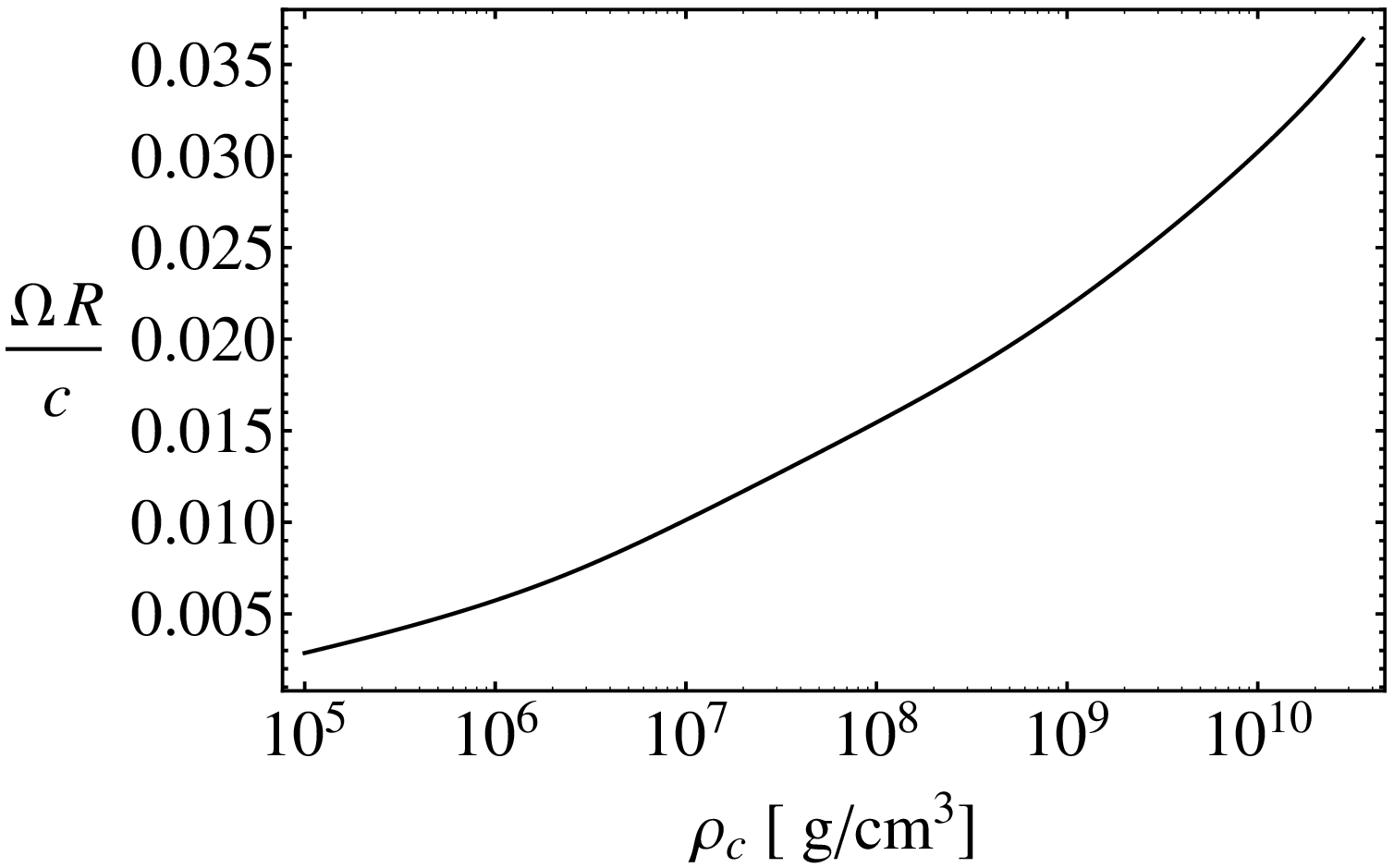} \includegraphics[width=0.48\hsize,clip]{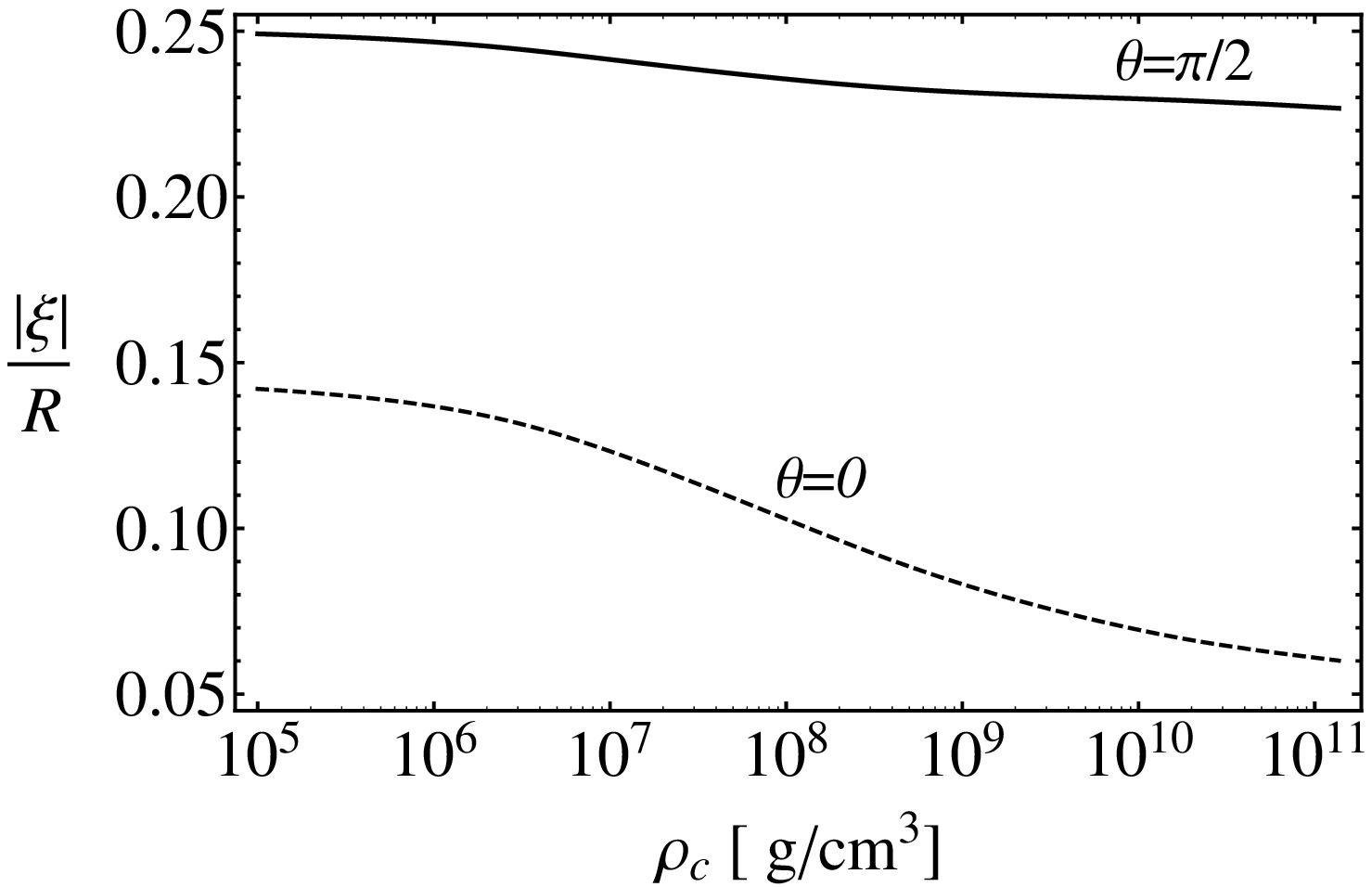}
\caption{Left panel: the rotation parameter normalized to the speed of light versus the central density. Right panel: the difference in the radial coordinate over the static radius versus the central density. The solid curve corresponds to the difference between equatorial ($\theta=\pi/2$) and static radii and the dashed curve corresponds to the difference between polar ($\theta=0$) and static radii.}\label{fig:xioverR}
\end{figure}

Now we turn to evaluate more specifically the deviations from the spherical symmetry. The expansion of the radial coordinate of a rotating configuration $r(R,\theta)$ in powers of the angular velocity is written as \citep{H1967}
\begin{equation}
r=R+\xi(R,\theta)+O(\Omega^4),
\end{equation}
where $\xi$ is the difference in the radial coordinate, $r$, between a point located at the polar angle $\theta$ on the surface of constant density $\rho(R)$ in the rotating configuration, and the point located at the same polar angle on the same constant density surface in the non-rotating configuration.
In the slow rotation regime, the fractional displacement of the surfaces of constant density due to the rotation have to be small, namely $\xi(R,\theta)/R\ll 1$, 
where $\xi(R,\theta)=\xi_0(R)+\xi_2(R)P_2(\cos\theta)$ and $\xi_0(R)$ and $\xi_2(R)$ are function of $R$ proportional to $\Omega^2$. On the right panel of Fig.~\ref{fig:xioverR} the difference in the radial coordinate over static radius versus the central density is shown. Here we see the same tendency as in the case of the eccentricity, that these differences are decreasing with an increasing central density. On the left panel the rotation parameter $\Omega R/c$ versus the central density is shown. Here, with an increasing central density the rotation parameter increases. Thus, for higher densities the system becomes less oblate, smaller in size with a larger rotation parameter i.e.~higher angular velocity.

In order to estimate the accuracy of the slow rotation approximation for RWDs, based on the above results, it is useful to compare all the above numbers with the known results for NSs. For instance, we notice that in NSs $\Omega R/c\sim10^{-1}$, $\xi(R,0)/R\sim10^{-2}$ and $\xi(R,\pi/2)/R\sim10^{-1}$ \citep[see e.g.][]{berti2005}, to be compared with the corresponding values of RWDs shown in Fig.~\ref{fig:xioverR}, $\Omega R/c\lesssim 10^{-2}$, $\xi(R,0)/R\sim10^{-2}$ and $\xi(R,\pi/2)/R\sim10^{-1}$. \cite{1992ApJ...390..541W} calculate the accuracy of the Hartle's second order approximation and found that the mass of maximally rotating NSs is accurate within an error $\lesssim 4$\%; \cite{benhar2005} found that the inclusion of third order expansion $\Omega^3$ improved the mass-shedding limit numerical values in less than 1\% for NSs obeying different EOS. On the other-hand, it is known that the ratio $T/|W|$ in the case of NSs is as large as $\sim 0.1$ in the Keplerian sequence (see e.g.~Tables 1--5 of \cite{berti2004}). Since RWDs have $T/|W|$ and $\Omega R/c$ smaller than NSs, and $\delta R/R=\xi/R$ at least of the same order (see left panel of Fig.~\ref{fig:ToverW}), we expect that the description of the strucure of RWDs up to the mass-shedding limit within the Hartle's approach to have at least the same accuracy as in the case of NSs.

\end{document}